\documentclass[dvipdfmx]{ptephy_v1}

\preprintnumber{XXXX-XXXX} 
\usepackage{graphicx}
\usepackage{url}

\newcommand{\amu}{\ensuremath{a_{\mu}}}

\newcommand{\mum}{$\mathrm{Mu^{-}}$}
\newcommand\rev[1]{#1}
\newcommand\revr[1]{#1}
\newcommand\revc[1]{#1}

\begin{document}

\title{First muon acceleration and muon linear accelerator for measuring the muon anomalous magnetic moment and electric dipole moment}

\author{Masashi Otani}
\affil{High Energy Accelerator Research Organization (KEK), Ibaraki 319-1106, Japan \email{masashio@post.kek.jp}}





\begin{abstract}%
Muon acceleration using a radio-frequency accelerator was recently demonstrated for the first time. 
Measurement of the muon anomalous magnetic moment and electric dipole moment at 
Japan Proton Accelerator Research Complex is the first experiment using accelerated muon beams, and construction will begin soon. 
The radio-frequency accelerator used in the experiment and the first muon acceleration are described in this paper. 
\end{abstract}

\subjectindex{C07, C31, G07}

\maketitle

\section{Introduction}
A remarkable development of accelerator science from the beginning of the 20th century to the present 
has enabled the use of various quantum beams, such as electron, proton, and ion beams, and has opened the way to greater human understanding, 
including the discovery of the Higgs boson and quantum beam therapy etc. 
The acceleration of muons using a radio-frequency accelerator was recently demonstrated for the first time~\cite{bib:Bae18}, 
and has opened a new era of accelerator science using accelerated muon beams. 
\par
A muon is an elementary particle similar to an electron, with an electric charge of  $-e$ and a spin of $\frac{1}{2}$, 
but with a mass 200-times heavier. 
After the successful generation of muons using a proton driver half a century after their discovery during the study of cosmic-rays~\cite{bib:And37, bib:Mic75}, 
muons are now widely used in various scientific fields.
In recent years, the demand for muon acceleration has been increasing in many different areas. 
For example, a muon collider, where muons are accelerated to a high energy for colliding, is one of the future plans of particle physics~\cite{bib:Man19, bib:Bos18}. 
In the material and life sciences, one promising
application of muon acceleration is in the construction of
a transmission muon microscope~\cite{bib:tmm}.
If the muons can be cooled to thermal temperature and subsequently re-accelerated,
transmission muon microscopes will be
realized.
\rev{Among the future programs, a new experiment (E34 experiment) in the Materials and Life Science Experimental Facility (MLF) of the Japan Proton Accelerator Research Complex (J-PARC) 
is planning to measure the muon anomalous magnetic moment (\amu) and search for the electric dipole moment (EDM) as a pioneer in muon acceleration~\cite{bib:Abe20}.}
\par
Although the discovery of the Higgs boson using the Large Hadron Collider (LHC) has established the Standard Model (SM) 
as a successful description of particle interactions, 
we are still confronted with many problems that can be solved only through experimental clues. 
One of the most interesting clues is the precise measurement of \amu, which has paved the way for understanding the nature of elementary particles through the quantum effects. 
In a series of three experiments at \rev{European Organization for Nuclear Research (CERN)}~\cite{bib:Cha62, bib:Bai72, bib:Bai79} and an experiment at Brookhaven National Laboratory (BNL)~\cite{bib:Ben06}, 
successive improvements in the accuracy of the measurements had enabled a deeper understanding of the SM and 
there is a large discrepancy between measurements and predictions of the SM. 
After many years of scrutiny, challenging calculations and corrections were conducted by stimulated theorists 
and still 3.7 sigma deviation remains~\cite{bib:Aoyama20}. 
This discrepancy should be addressed by new measurements. 
A new experiment at Fermi National Accelerator Laboratory (FNAL E989~\cite{bib:Gra15}) is currently being conducted, with the storage ring from the BNL experiment being reused.
\rev{They published their first result in early FY2021~\cite{bib:abi21}. 
The result is consistent to that of the previous BNL experiment and 
the the tension between experiment and the SM calculation becomes 4.2 sigma. 
It strengthens the importance for confirming the tension with with independent measurements different from BNL and FNAL. }
\par
As described above, continuous studies have continued improving the accuracy of \amu. 
In particular, the improvement of the beam has been one of the driving forces moving the measurements forward. 
Our ancestors at CERN, BNL, and FNAL have been struggling with beam-related uncertainties in their measurements 
because they use muons obtained directly from a pion decay with a large emittance. 
The J-PARC E34 experiment~\cite{bib:Abe20} aims to measure \amu~ with a precision of 0.1 ppm using an unprecedentedly low-emittance muon beam realized by the acceleration of thermal muons. 
\par
The reminder of this paper is organized as follows. 
In \revr{Section}~\ref{sec:e34}, the J-PARC E34 experiment, particularly the radio-frequency linear accelerator (linac) dedicated to muons is described. 
Section~\ref{sec:muacc} describes the first demonstration of muon acceleration. 
A summary and outlook are shown in \revr{Section}~\ref{sec:sum}. 
\section{Muon linac for the J-PARC E34 experiment}\label{sec:e34}
In this section, details of the muon linac will be described after successive descriptions explaining an overview of the J-PARC E34 experiment.
\par
\begin{figure}[!h]
\centering\includegraphics[width=0.95\textwidth]{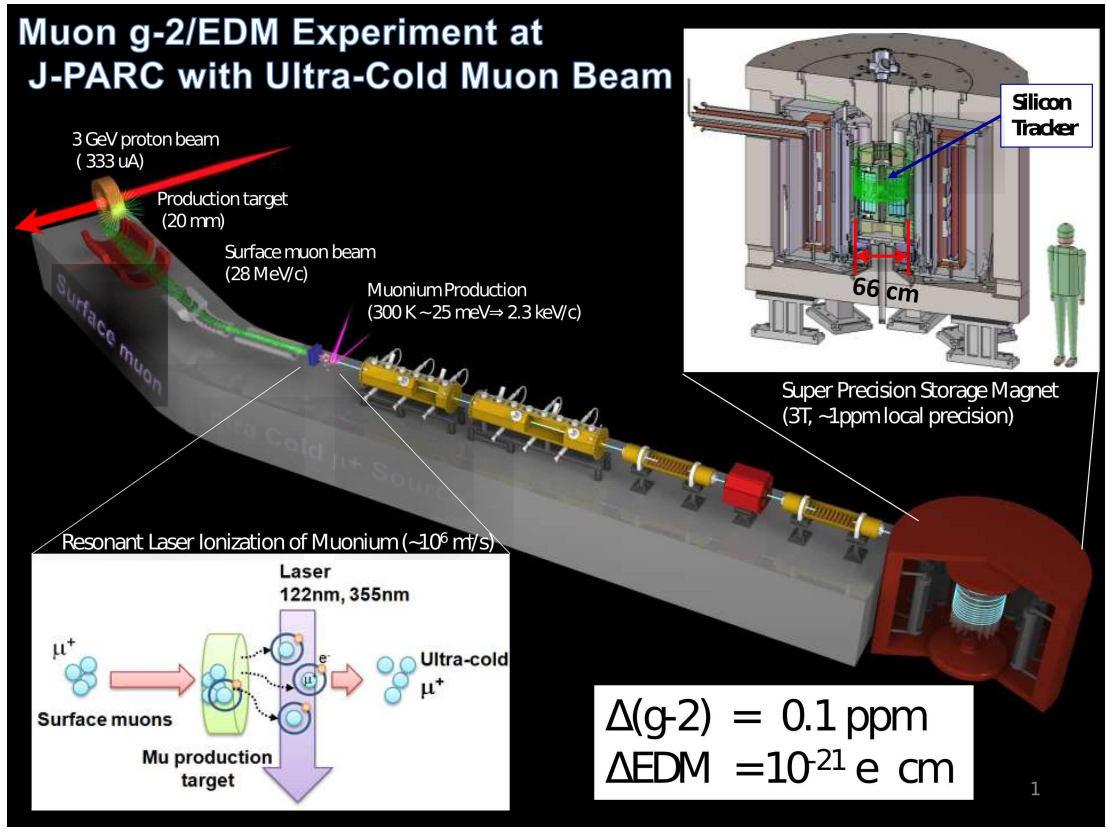}
\caption{Overview of the J-PARC E34 experiment. }
\label{fig:e34}
\end{figure}

The J-PARC E34 experiment aims to measure \amu~ with a precision of 0.1 ppm and search for the EDM with a sensitivity 
of approximately $10^{-21}$~$e\cdot$cm using a low-emittance muon beam realized by an accelerated thermal muon beam. 
\rev{The total emittance in the transverse direction is required to be 1.5$\pi$~mm~mrad to realize the measurement 
with a 3 T compact MRI-type magnet with sufficient injection efficiency. 
The cyclotron radius with this magnet is 333~mm that is about a factor of 20 smaller than that for the BNL and FNAL experiments. 
Because of the high uniformity of the magnetic field in the muon storage region, the uncertainty due to the field uniformity is much smaller than in the BNL and FNAL experiments. }
Figure~\ref{fig:e34} shows the experimental setup. 
The pulsed high-power primary proton beam generates secondary surface muons produced by $\pi^{+}$ decay 
near the surface of the production target~\cite{bib:Kawamura18}. 
The produced surface muons are extracted and thermalized to form muoniums \rev{(bound state made up of a positive muon and an electron)}, which are then emitted into vacuum region adjacent to the muonium production target~\cite{bib:Bak13, bib:Beer14, bib:Bea20}.
The paired electron in the muonium is knocked out by a laser, and thermal muon (3 keV/c) is generated. 
After acceleration to 300~MeV/c, the muon beam has an extremely low emittance such that it can be injected and stored in a high precision compact storage magnet~\cite{bib:Iinuma16, bib:Abe18}, 
where the time dependence of the positrons from the muon decay is measured for the measurement of the anomalous spin precession~\cite{bib:Aoyagi20, bib:Kishishita20}. 
\par
\par
\subsection{Overview of muon linac}
Because muons have a finite lifetime of approximately \rev{2.2}~$\mu$s, they need to be accelerated faster to avoid decay losses to obtain the necessary experimental statistics. 
From this perspective, a linac is one of the best options for muon acceleration. 
\par
\begin{figure}[!h]
\centering\includegraphics[width=0.95\textwidth]{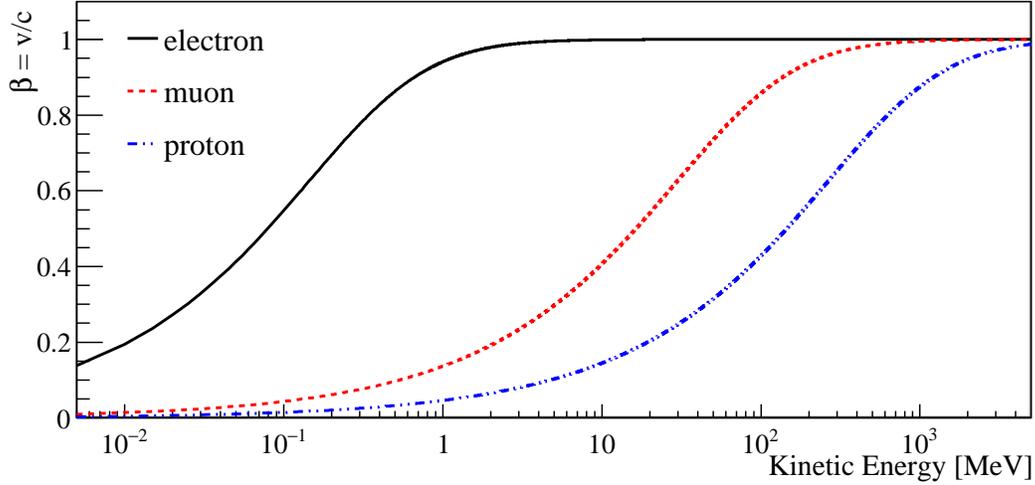}
\caption{Particle velocity as a function of kinetic energy for electron (black line), muon (red hatched line), and proton (blue dot-dash line)}
\label{fig:velocity}
\end{figure}

Owing to its intermediate mass between that of protons and electrons, 
the change in velocity upon acceleration is slower than that of electrons, as shown in Fig.~\ref{fig:velocity}. 
For this reason, both proton and electron linac technologies will be used to accelerate the muon to near the speed of light. 
\par
Accelerators should be designed appropriately to reduce costs. 
To reduce the expenses, a spare radio-frequency quadrupole (RFQ) of the J-PARC linac~\cite{bib:Kondo13} will be used as 
a first stage acceleration. 
It operates with a resonant frequency of 324~MHz. 
The L-band high-power klystron developed for the KEKB linac upgrade~\cite{bib:Kubosaki11} will be used as the RF power supply for the acceleration cavity in the high-energy section to achieve a further cost reduction.
\par
\begin{figure}[!h]
\centering\includegraphics[width=0.98\textwidth]{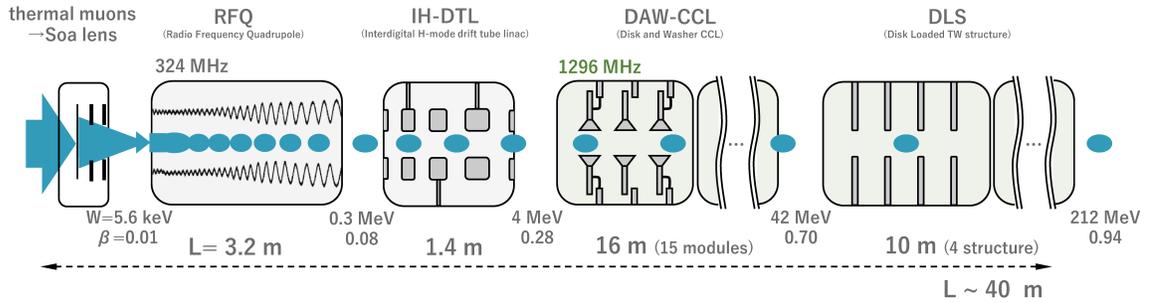}
\caption{Configuration of the muon linac. \rev{In this figure, $W$ is kinetic energy and $\beta$ is the ratio of the muon velocity to the speed of light}.}
\label{fig:mulinac}
\end{figure}

A schematic of the muon linac is shown in Fig.~\ref{fig:mulinac}. 
The RFQ bunches and accelerates the muons to 0.3~MeV after the initial electrostatic acceleration~\cite{bib:Kondo15}. 
After the RFQ, an inter-digital H-mode drift tube linac (IH-DTL) is employed during the particle velocity $\beta=0.08$ to $0.28$ (4~MeV)~\cite{bib:Otani16}. 
After the muon is accelerated to $\beta=0.28$, a disk-and-washer (DAW) type coupled cavity linac (CCL) with an operational frequency of 1296~MHz is employed~\cite{bib:Otani19daw}. 
Because the $\beta$ variation is modest in the high-$\beta$ region, to realize a sufficiently
short distance, the design emphasis has been shifted to achieving a high accelerating gradient. 
A disk-loaded structure (DLS) traveling-wave linac is used when $\beta$ is greater than $0.7$ (42~MeV)~\cite{bib:Kondo17}. 
The details of each acceleration cavity are described below. 
\subsection{Initial electrostatic acceleration}
In front of the RFQ, there is an electrostatic lens called a Soa lens~\cite{bib:Can86}, which accelerates and extracts the thermal muons. 
\par
\begin{figure}[!h]
\centering\includegraphics[width=4in]{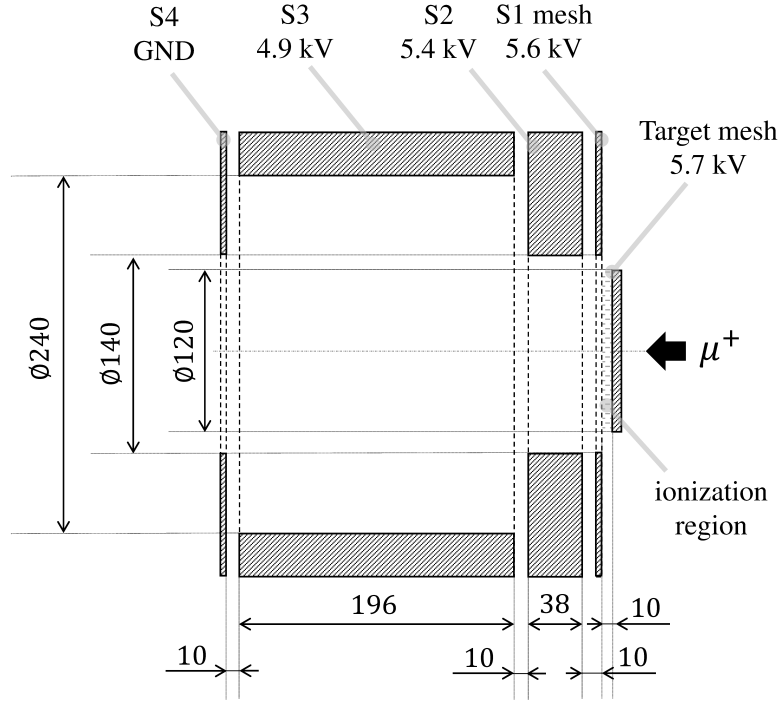}
\caption{Schematic of the SOA lens. \rev{Lengths in the figure are expressed in millimeters.}}
\label{fig:soa}
\end{figure}

Figure~\ref{fig:soa} shows a schematic view of the Soa lens. 
The Soa lens consists of two mesh electrodes and three cylindrical electrodes. 
The first mesh electrode (target mesh) covers the downstream surface of the silica aerogel target. 
The laser ionization region is between the target mesh and second mesh electrode (S1). 
The voltage applied to the target mesh and S1 is set to 5.7 and 5.6~keV, respectively, corresponding to the input energy of the RFQ. 
The dimensions of the electrodes are designed to cover the primary surface muons (rms size of 31 mm and 14 mm in the horizontal and vertical directions, respectively~\cite{bib:Otani18_surfmu}) 
and to provide a sufficient extraction efficiency for ultra slow muons. 
The voltage applied to other electrodes (S2, S3 and S4) is determined using the simulation described below 
such that the phase space of the beam matched with the design acceptance of the RFQ and a high transmission efficiency is obtained.
\par
Input muon distributions are estimated using a simulation based on measurements. 
The simulation for a surface muon beamline (MLF H-line) is constructed using the g4beamline~\cite{bib:g4bl, bib:Kawamura18}. 
The absolute number of surface muons is normalized by measurements from another beamline (MLF D-line~\cite{bib:Stra10}) that uses the same muon production target in J-PARC MLF.
The stopping distribution inside the silica aerogel is estimated using GEANT4~\cite{bib:g4}. 
The simulation for muonium diffusion is developed based on a three-dimensional random walk. 
The simulation parameters of the thermal temperature and the diffusion constant were determined from our measurements at \rev{Canada's particle accelerator centre (TRIUMF)}~\cite{bib:Beer14}. 
The muon trajectories are simulated using the GEANT4 simulation where the electrostaic field of the Soa lens calculated OPERA~\cite{bib:opera} is implemented.  
\par
\begin{figure}[!h]
\centering\includegraphics[width=5in]{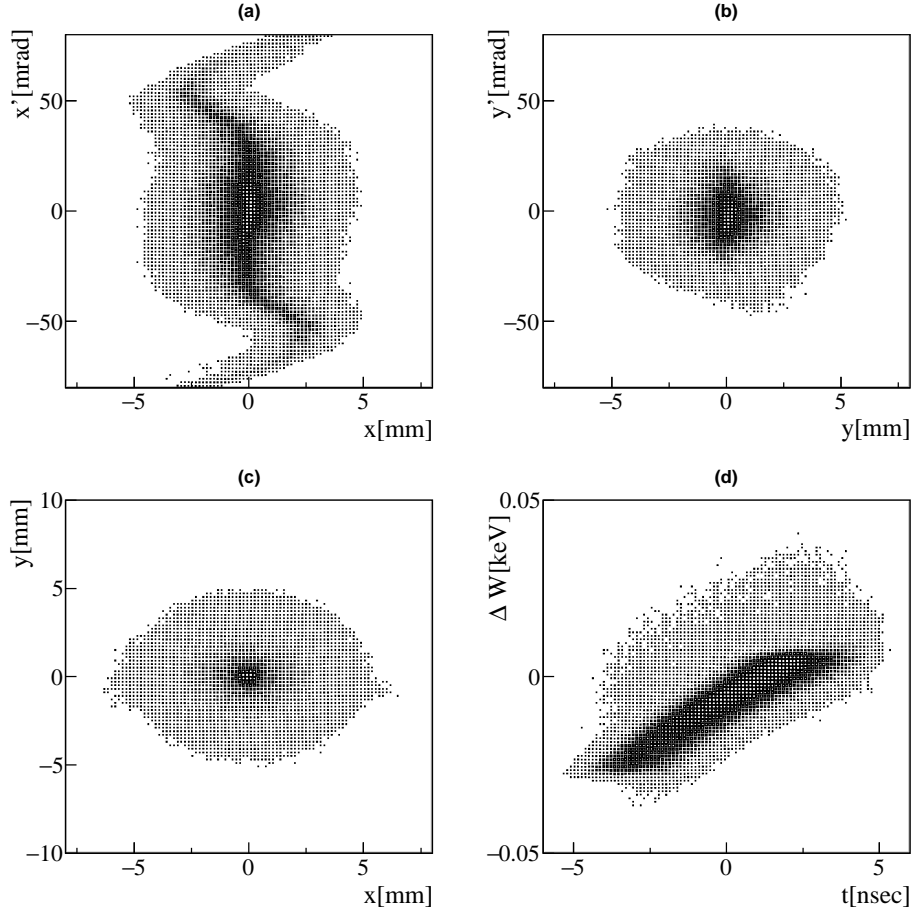}
\caption{Phase space distributions at the entrance of the RFQ: (a) Horizontal divergence angle x' 
vs. x, (b) the vertical divergence angle y' vs. y, (c) y vs. x, and (d) $\Delta W$ vs. time}
\label{fig:soa_xyps}
\end{figure}

Figure~\ref{fig:soa_xyps} shows the transverse phase space distributions at the entrance of the RFQ. 
The difference between the horizontal and vertical directions is due to the difference in the primary surface muon distribution at the entrance of the silica aerogel. 
Owing to the spatial distribution of the muoniums in the laser ionization region, the muons are distributed over time with a full width of approximately 10~ns. 
The transmission efficiency in the Soa lens is estimated to be 72\%, including a 17\% decay loss. 
\rev{Because the structure of the mesh electrodes (target mesh and S1) is not implemented in the  GEANT4 simulation, 
the transmission efficiency of the mesh electrode was estimated to be 78\% based on the products of the aperture ratio of the two meshes. 
In total, the efficiency is 56\% in the initial electrostatic acceleration. }
\par
\subsection{RFQ}
After the initial acceleration by the Soa lens, the RFQ accelerates the muons to 340~keV. 
In addition to the acceleration, the RFQ bunches the muons at a frequency of 324 MHz. 
\par
\rev{
The principles of the RFQ were first invented in 1969~\cite{bib:Kap70_1, bib:Kap70_2}
and \revr{they proved} at Los Alamos National Laboratory (LANL) in 1980~\cite{bib:Sto81}. 
The RFQ consists of a four electrodes excited with quadrupole-mode. 
The electrode is modulated longitudinally, generating the axial electric field. 
By changing modulation pattern gradually so that the synchronous phase is changed from $-90$ degree to higher, the beam bunching accomplished. 
Because the RFQ supply velocity-independent electric focusing, it has great advantage in a low-velocity part compared with conventional linacs 
that used velocity-dependent magnetic focusing. 
}
\par
Table~\ref{tbl:rfqpqr} lists the design parameters of the RFQ for the muon acceleration along with that for negative hydrogen ion (H$^-$). 
To accelerate muons using the spare of the J-PARC RFQ, which was originally developed for H$^-$ acceleration, 
the intervane voltage must be reduced to the mass ratio to match the particle velocity. 
As a result, the required power is reduced to the square of the mass ratio. 
The input and output energies are also scaled to the design velocity; 
the input and output $\beta$ are 0.01 and 0.08, respectively. 
\begin{table}[h!tbp]
   \centering
   \caption{Parameters of the RFQ~II for the H$^{-}$ and muon acceleration.}
   \label{}
   \begin{tabular}{lll}
       \hline
                                                   & H$^-$ & muon             \\
       \hline
           Frequency (MHz) & \multicolumn{2}{c}{324}    \\
           Number of cells                & \multicolumn{2}{c}{295}            \\
           Length (m)                       & \multicolumn{2}{c}{3.17}          \\
           Intervane voltage (kV)       & 82.9    & 9.3               \\
           Power (kW)                       & 330    &  4.2              \\
           Injection energy (keV)        & 50  & 5.6                \\
           Extraction energy (MeV)    & 3    & 0.34 \\
       \hline
   \end{tabular}
   \label{tbl:rfqpqr}
\end{table}
\par
\begin{figure}[!h]
\centering\includegraphics[width=5in]{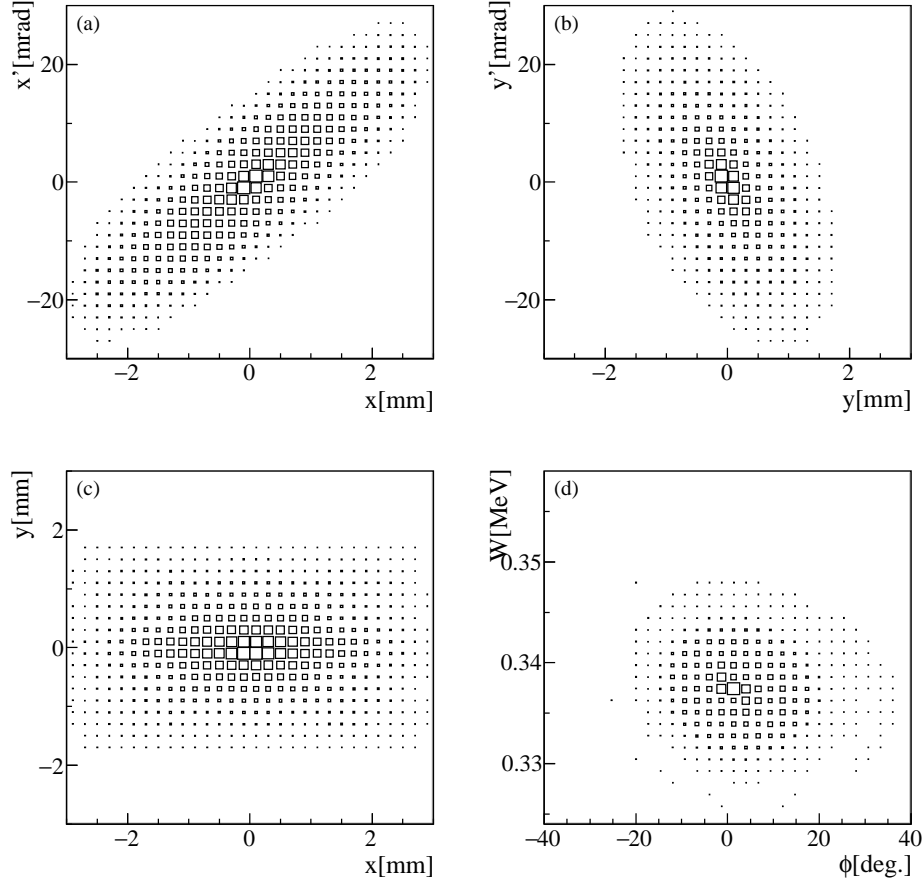}
\caption{Phase space distributions at the RFQ exit: (a) Horizontal divergence angle x' 
vs. x, (b) vertical divergence angle y' vs. y, (c) y vs. x, and (d) $\Delta W$ vs. \revc{rf phase ($=$ time)}}
\label{fig:rfqsim}
\end{figure}

In order to confirm that the RFQ can accelerate muons without any problems, 
particle simulations were performed using PARMTEQM~\cite{bib:parmteqm}. 
Figure~\ref{fig:rfqsim} shows the phase space distributions at the RFQ exit, and 
Table \ref{tbl:rfqsim} summarizes the input and output beam parameters. 
The horizontal and vertical normalized rms emittances at the RFQ exit are 0.30 and 0.17 $\pi$~mm~mrad, respectively.
The simulated transmission is 94.7\%. 
Since the transit time through one cell of the RFQ is half the resonant frequency, the transit time through the entire 
RFQ can be calculated as $\frac{1}{324 \times 10^6} \times \frac{1}{2} \times 295 = 455$ ns. 
Ignoring relativistic corrections, the muon lifetime is 2.2~$\mu$s, and the survival rate of muons passing through the RFQ is $\exp(-0.455/2.2) = 0.813$. 
Therefore, the total transmission is 77.0\%. 
\par
\begin{figure}[!h]
\centering\includegraphics[width=5in]{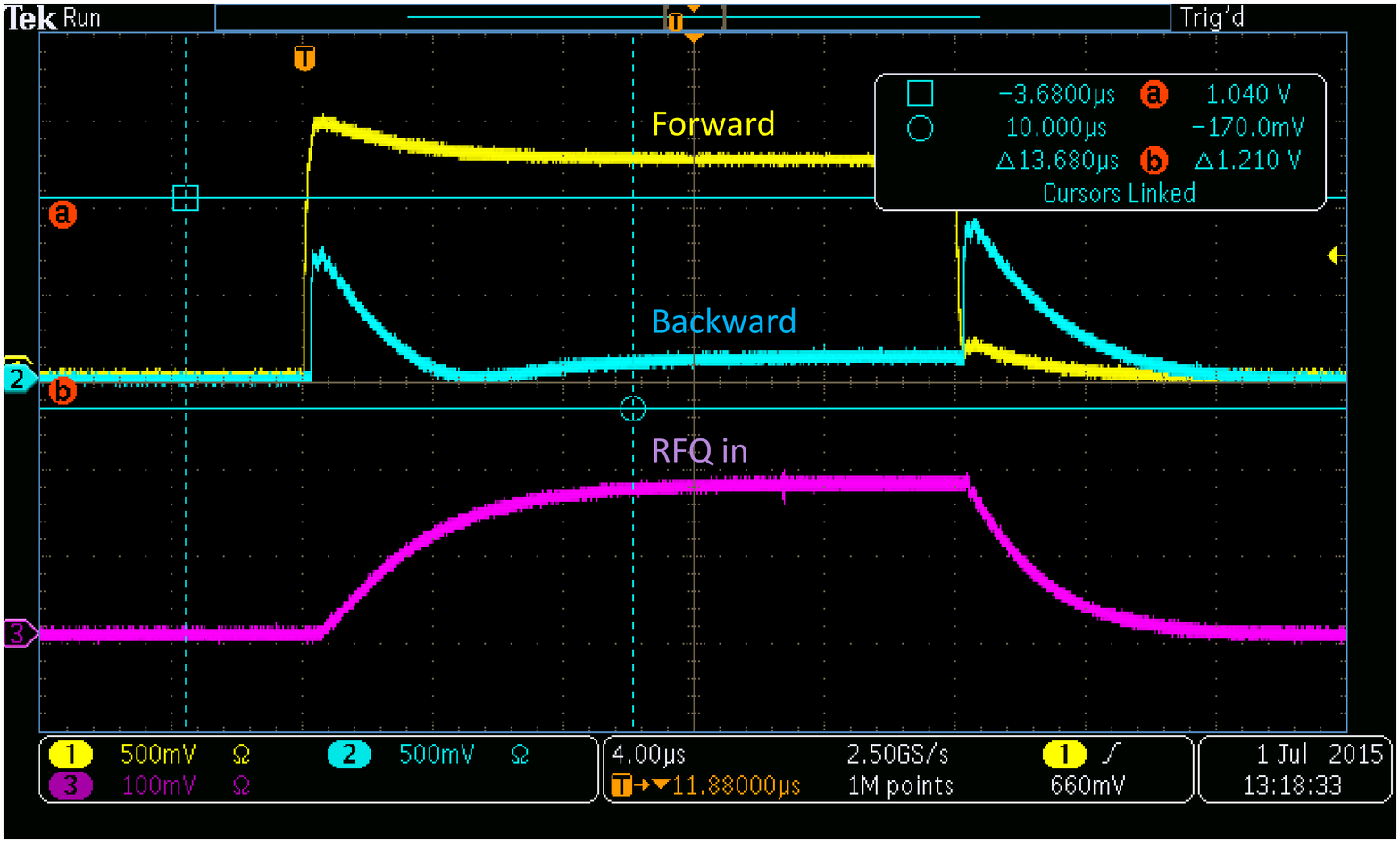}
\caption{Forward, reflection wave, and pick-up power in RFQ with nominal power of 5 kW }
\label{fig:rfq_offline}
\end{figure}

\begin{table}[h!tbp]
   \centering
   \caption{Input and output beam parameters of the \revc{beam} dynamics simulation of the RFQ. 
   \revc{In this table, $\alpha$ and $\beta$ are Twiss parameters. } }
   \begin{tabular}{lll}
       \hline
                                     & Input & Output                                       \\
       \hline
           $\alpha_x$                                                      & 0.32    & -1.51                                \\
           $\beta_x$ [mm/mrad]                                        & 0.041  & 0.21            \\
           $\varepsilon_x$ [$\pi$~mm~mrad, rms, normalized ]   & 0.376   & 0.297 \\
           $\alpha_y$                                                      & 0.092 & 0.606 \\
           $\beta_y$ [mm/mrad]                                        & 0.080    & 0.076                 \\
           $\varepsilon_y$ [$\pi$~mm~mrad, rms, normalized ]   & 0.106 & 0.167 \\
           $\alpha_z$                                                       & -  & 0.17                                 \\
           $\beta_z$[deg/MeV]                                         & - & 1360                        \\
           $\varepsilon_z$[$\pi$~MeV~deg, rms, normalized ]    & -  & 0.0381 \\
           Time width        & 10~ns (full width) &    -               \\
           Energy spread     & 0.00989~keV (rms) &  -                  \\
       \hline
           Transmission       & \multicolumn{2}{c}{ 94.7\%    }                            \\
           Transient time     & \multicolumn{2}{c}{ 455 ns   }                             \\
           Survival rate      &\multicolumn{2}{c}{ 81.3\%     }                           \\
           Transmission total & \multicolumn{2}{c}{ 77.0\%}                                \\
       \hline
   \end{tabular}
   \label{tbl:rfqsim}
\end{table}
The operation test of the RFQ has already been conducted~\cite{bib:Otani15_pasj}. 
The RFQ was powered on by a solid state amplifier at up to 6 kW and a 25 Hz repetition. 
Vacuuming is done with an ion pump and reaches $10^{-6}$ Pa. 
Figure~\ref{fig:rfq_offline} shows the forward, reflection, and pick-up power in the RFQ when 5~kW power is applied. 
There is a small reflection because the coupling was tuned to be overcoupled originally for a high current H$^-$ beam. 
A few hours of operation at 5~kW was successfully achieved without any problems, such as an RF failure from a spark. 
The background associated with the RF operation was measured using a micro-channel plate detector connected downstream of the RFQ. 
The detector count rates were consistent and negligibly small with and without the RF operation, within a statistical uncertainty range of 0.1~Hz. 
\par
In conclusion, we are ready for a muon acceleration using the RFQ.
\subsection{IH-DTL}
After the initial acceleration and bunching by the RFQ, the IH-DTL accelerates the muons to 4.3~MeV. 
\par
The IH-DTL has alternating drift tubes up and down through the stem allowing the TE11 mode to be used for acceleration. 
\rev{The IH-DTL was first proposed in Japan in 1949~\cite{bib:Mor49}. 
\revr{Although} there were several efforts for this inventions~\cite{bib:Ble56, bib:Zei62, bib:Pot69}, an IH-DTL was realized after a quarter century as the heavy ion post-accelerator at the Munich tandem accelerator~\cite{bib:Nol79}. }
Compared to the conventional Arvaretz DTL, a higher acceleration efficiency is achieved, particularly within the range of $\beta=0.1$--$0.2$~\cite{bib:Rat05}. 
In recent IH-DTLs, the alternating phase focusing (APF) method~\cite{bib:Good53, bib:Min99}, which was successfully implemented in the Hadron therapy machine~\cite{bib:Iwata06}, enables simultaneous focusing in the horizontal and vertical directions using only electric fields, resulting in a higher acceleration efficiency. 
Owing to the small focusing strength in the APF method, there is a limit to its use in high current machines, 
but it can be applied in the muon linac, which has a small current. 
\par
The IH-DTL with the APF method for accelerating muons was designed using several types of simulation software~\cite{bib:Otani16}. 
The arrangement of the drift tubes and acceleration gaps, i.e., the so-called longitudinal beam dynamics design, should be determined according to the evolution of the beam velocity, 
which is determined by the energy gain at each gap.
Unlike conventional accelerating cavities where transverse focusing is performed conducted using additional elements such as quadrupoles, 
an APF cavity requires simultaneous non-independent longitudinal and transverse beam dynamics designs. 
This beam dynamics design was performed based on linacsapf~\cite{bib:jameson14}, where the 
beam dynamics is calculated using the so-called {\it drift-kick-drift} approximation. 
To achieve less emittance growth and a better transmission, a nonlinear optimization is applied to the synchronous phase array. 
The cavity was designed using CST-MW Studio~\cite{bib:cst} based on the beam dynamics design. 
In order to mitigate the peak electric field and obtain a higher efficiency, the dimensions of the drift tube were changed from \cite{bib:Otani16}. 
Table~\ref{tbl:ihdesign} summarizes the basic parameters of the IH-DTL cavity. 
\par
\begin{figure}[!h]
\centering\includegraphics[width=5in]{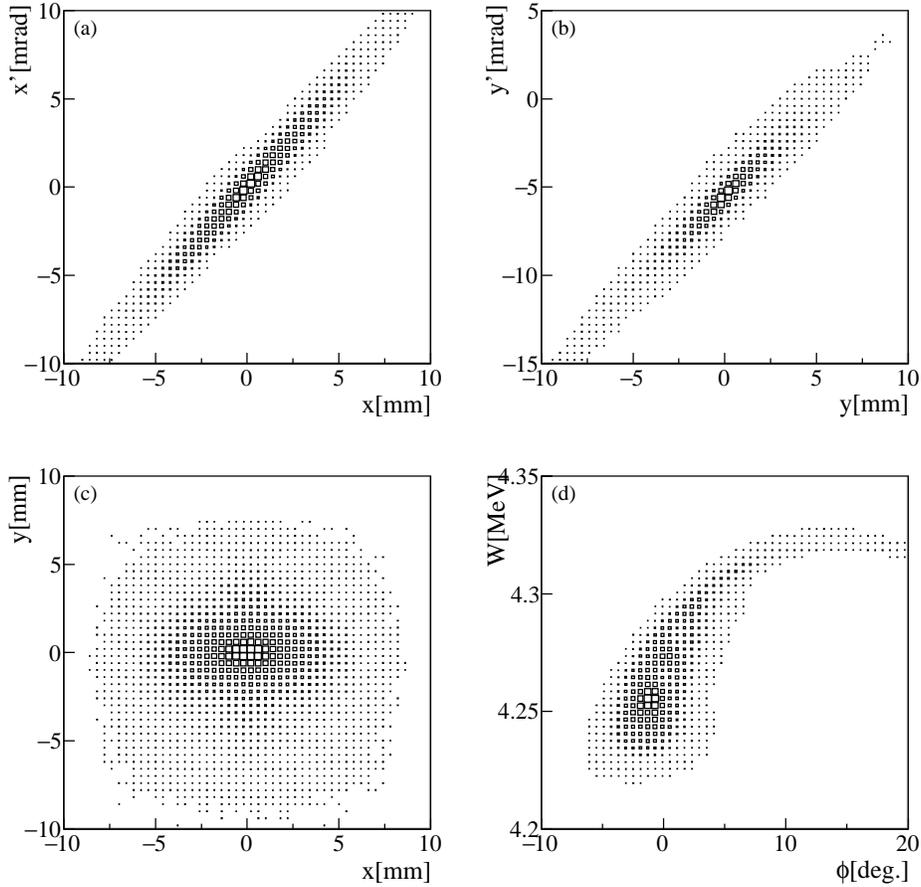}
\caption{Phase space distributions at the IH-DTL exit: (a) Horizontal divergence angle x' 
vs. x, (b) vertical divergence angle y' vs. y, (c) y vs. x, and (d) $\Delta W$ vs. \revc{rf phase ($=$ time)}}
\label{fig:ihsim}
\end{figure}

\begin{table}[h!tbp]
   \centering
   \caption{Parameters of the IH-DTL cavity. Because the frequency is tuned to 324~MHz with conductive tuners, the frequency shown here is slightly smaller than 324~MHz. 
In this table, \rev{$E_{surf.}$ is the surface electric field and} $E_{kp}$ is the Kilpatrick limit at 324~MHz (17.8~MV/m).}
   \begin{tabular}{lll}
       \hline
           Parameters & Values                                 \\
       \hline
           Frequency (MHz)          &     323.3                           \\
           Number of cells           &      16                           \\
           Length (m)                  &     1.32                            \\
           Synchronous phase [deg.]      &      $-44$--$+48$                           \\
           Unloaded $Q$              &      $1.04\times10^{4}$                           \\
           Power (kW)                  &      322                           \\
           Peak $E_{surf.}$ (MV/m) &     35.3 ($\sim1.99 E_{kp}$)                            \\
       \hline
   \end{tabular}
   \label{tbl:ihdesign}
\end{table}
The input distribution was given from the upstream RFQ simulation. 
The input was used for the beam dynamics simulation after simulation of the beam transport line to match the beam to the IH-DTL. 
The beam dynamics simulation is performed using GPT~\cite{bib:gpt}. 
Figure~\ref{fig:ihsim} shows the phase space distributions at the IH-DTL exit, and 
Table \ref{tbl:ihsim} summarizes the output beam parameters. 
The vertical emittance growth is larger than the horizontal one, 
which is due to the vertical meandering track caused by the vertical electric field and horizontal magnetic fields of the first and last cells, which is inevitable in IH-DTL structures. 
The transmission is \rev{99.97}\% and it is sufficiently large. 
The beam transit time \rev{$t_{\mathrm{trans.}}$} is 25~ns and the muon survival rate is calculated to be $\exp(t_{\mathrm{tran.}}/\tau_{\mu}\overline{\gamma})=98.9$\%, where the average Lorentz factor during acceleration is labeled $\overline{\gamma}$ \rev{and $\tau_{\mu}$ is the muon lifetime (2.2~$\mu$s)}. 
\par
\begin{table}[h!tbp]
   \centering
   \caption{Output beam parameters of the IH-DTL. \revc{In this table, $\alpha$ and $\beta$ are Twiss parameters. } }
   \begin{tabular}{ll}
       \hline
                                     & Output                                       \\
       \hline
           $\alpha_x$                                                      & -4.3                                \\
           $\beta_x$ [mm/mrad]                                     &  0.38            \\
           $\varepsilon_x$ [$\pi$~mm~mrad, rms, normalized ]   & 0.316 \\
           $\alpha_y$        & -3.4 \\
           $\beta_y$ [mm/mrad]         & 0.33                 \\
           $\varepsilon_y$ [$\pi$~mm~mrad, rms, normalized ]   & 0.190 \\
           $\alpha_z$        & -1.0                              \\
           $\beta_z$[deg/MeV]         & 174                        \\
           $\varepsilon_z$[$\pi$~MeV~deg, rms, normalized ]  & 0.0274 \\
       \hline
           Transmission       & 99.97\%                                \\
           Transient time     & 25 ns                                \\
           Survival rate      & 98.9\%                               \\
           Transmission total & 98.9\%                                \\
       \hline
   \end{tabular}
   \label{tbl:ihsim}
\end{table}
An error study of the beam dynamics was conducted by assuming several possible cases~\cite{bib:Otani16pasj}. 
The error in the on-axis electric field owing to a fabrication error is estimated to be approximately 2\% using the CST-MW Studio
assuming a general fabrication error of 100~$\mu$m. 
The emittance growth due to the fabrication errors is estimated to be 10\%, which still satisfies the requirement. 
It also shows that the error field can be controlled by some movable conductive tuners installed on the cavity side wall. 
\par
\begin{figure}[!h]
\centering\includegraphics[width=0.95\textwidth]{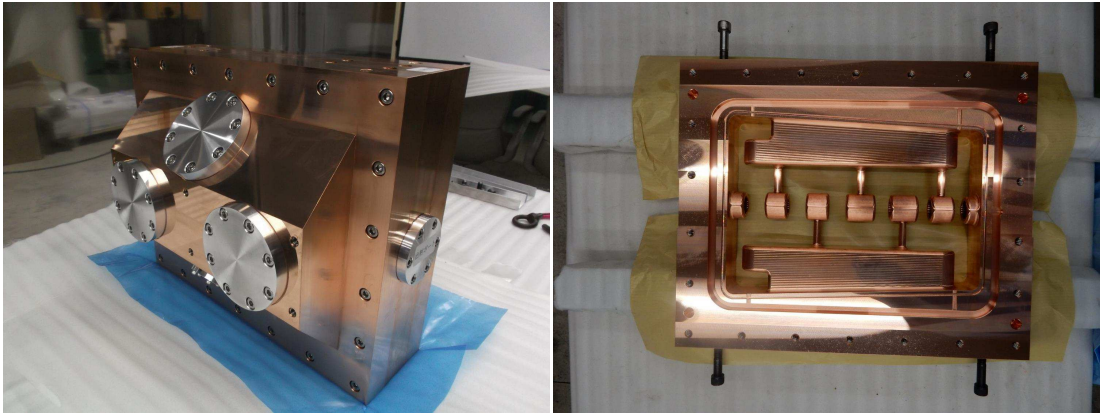}
\caption{Prototype of the IH-DTL: (Left) overall structure, (Right) center plate, where the five drift tubes are implemented. }
\label{fig:ihproto}
\end{figure}

A prototype of the IH-DTL was fabricated (Fig.~\ref{fig:ihproto}) to confirm the design and evaluate the performance~\cite{bib:Nakazawa19}. 
The prototype was a three-piece design with two semi-cylindrical shells attached to the center frame where the drift tubes are mounted.
It has the first five drift tubes of the actual design, and the total length is approximately 0.5 m. 
The resonant frequency and unloaded quality factor ($Q_0$) were measured and are consistent with the \rev{CST-MW Studio} calculation at 0.1\% and 11\%, respectively. 
The on-axis electric field distribution was measured using the bead-pull method~\cite{bib:bpm} and agreed with the \rev{CST-MW Studio} calculation within 3\% uncertainty. 
The emittance growth owing to the field error is estimated to be 3\%, which is negligible. 
A high power coupler was fabricated and a high power test will soon be conducted. 
Based on the experience with the prototype, we completed a detailed design of the actual IH-DTL and will soon start production. 
\subsection{DAW-CCL}
After the IH-DTL, the DAW-CCL accelerates the muons to 40~MeV. 
\par
A DAW-CCL is a cylindrical cavity with conductive washers placed across a disk-shaped disc. 
\rev{A DAW-CCL was first proposed in early 1970s~\cite{bib:and72, bib:Mur72} and full scale cavity was demonstrated in late 1970s~\cite{bib:and76}. 
The first DAW-CCL was operated in a proton and H$^-$ linac at the Moscow meson factory~\cite{bib:Esi88}. 
In Japan, DAW-CCLs were developed in KEK and Kyoto for electron~\cite{bib:Ina86, bib:Iwa94}. }
DAW-CCLs have advantages over other CCLs, such as a higher shunt impedance and higher coupling between the accelerating and coupling cells. 
\par
The DAW-CCL cavity is designed using several types of softwares~\cite{bib:Otani19daw}. 
As a first step, the two-dimensional dimensions are optimized using \rev{Poisson Superfish}~\cite{bib:sf} for a higher acceleration efficiency 
and lower peak-to-average value ($E_{\mathrm{max}}/E_{0}$) in satisfying the confluence condition. 
After the two-dimensional calculation, the three dimensional calculations including the washer support are conducted using CST-MW Studio. 
The bi-periodic L-support~\cite{bib:ao00}, 
in which a washer is fixed  by two supports and a pair of supports are located azimuthally 90 degrees apart from the adjacent supports,  
is adopted among several support structures because perturbation to the acceleration mode can be minimized by adjusting the support structure. 
Finally, the dispersion curve is investigated to check whether an unfavorable mode exists around the operation frequency. 
\par
\begin{figure}[!htb]
   \centering
    \includegraphics*[width=0.95\textwidth]{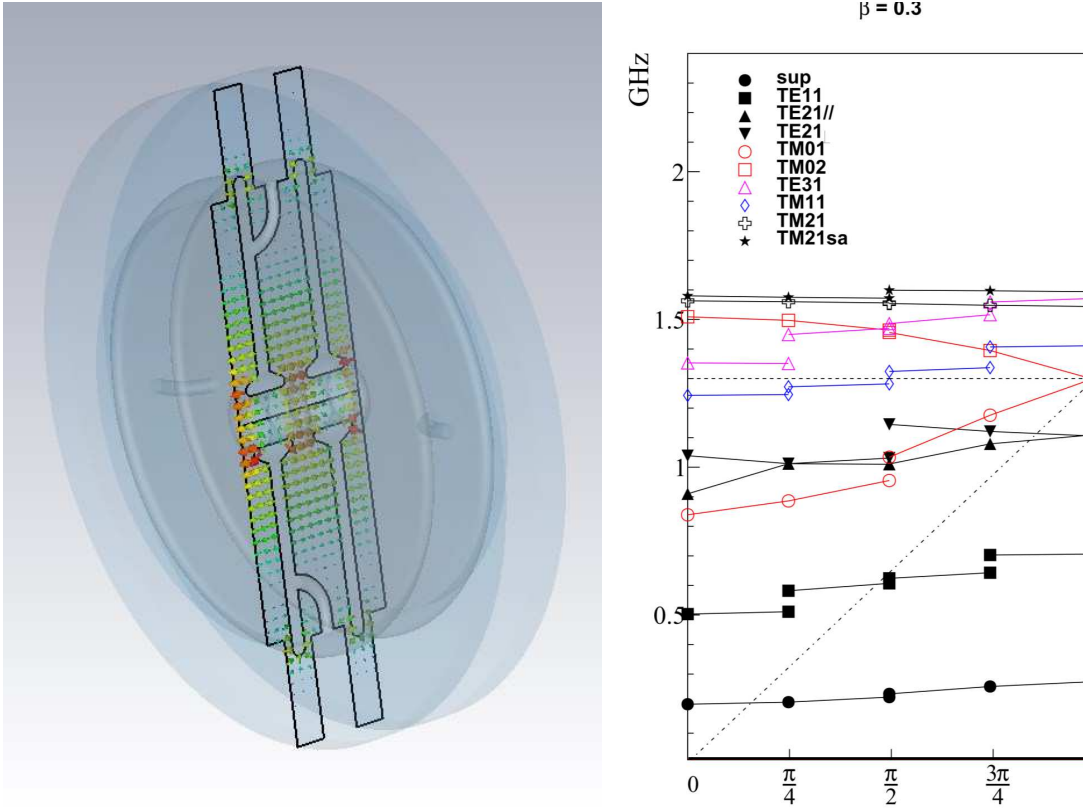}
   \caption{(Left) Three-dimensional model of the DAW-CCL. 
The electric field of the acceleration mode in the two-cell model is shown by the colored arrows. 
 (Right) Dispersion curve with optimized cavity in $\beta=0.3$ calculated using CST-MW Studio.
\rev{The dotted line represents target frequency (1.3~GHz) and the dashed line corresponds to the phase velocity of the muons. }
\revr{The sup mode is associated with the L-support.}
 }
   \label{fig:dawmodel}
\end{figure}

Figure~\ref{fig:dawmodel} shows the three-dimensional model (left) and the dispersion curve (right) of the designed cavity for $\beta=0.3$. 
TM02 and TM01 are accelerating and coupling mode, respectively. 
Because of the bi-periodic structure, some stop bands appear in $\pi/2$. 
Although the TM11 mode is near the operational frequency, the cavity is tuned during the optimization process such that the operational frequencies sit within the stop band at $\pi/2$. 
Although the dipole mode passband TE11 crosses the line where the phase velocity matches the speed of the muons, it is not considered to be a problem because the muon beam current is negligible and the transverse kick owing to this mode is estimated to be much smaller than the requirement. 
Table~\ref{tbl:dawcavity} summarizes the cavity parameters for $\beta=0.3, 0.4, 0.5,$ and $0.6$. 
\par
\begin{table}[!htb]
   \centering
   \caption{Parameters of the DAW-CCL cavity for $\beta=0.6, 0.5, 0.4,$ and $0.3$. For the simplicity, the resonant frequency is designed to be 1.3~GHz. 
   \rev{In this Table, L is the cell length, 
   $\mathrm{f_a}$ is the resonant frequency of the accelerating mode, 
   $\mathrm{f_c}$ is the resonant frequency of the coupling mode, 
   ZTT is the effective shunt impedance per unit length, 
   $E_{\mathrm{max}}/E_{0}$ is the peak to average axial electric-field, and 
   $\lambda$ is the resonant wavelength. }}
   \begin{tabular}{lrrrr}
       \hline
       	Parameters & \multicolumn{4}{c} {Values}   \\
       \hline
          $\beta$                                                         &    0.6            &    0.5         &    0.4        &   0.3          \\
          $\mathrm{L}$                                           &    \multicolumn{4}{c} {$\beta\lambda/4$}        \\
          $\mathrm{f_{a}[GHz]}$                             &     1.300         &    1.300         &     1.299        & 1.301      \\
          $\mathrm{f_{c}[GHz]}$                             &     1.299         &    1.301         &     1.302        & 1.301      \\
          $\mathrm{ZTT[M\Omega/m]}$                  &     57.8         &    46.3            &     33.8           & 18.0     \\
          Transit time factor                                 &     0.84         &    0.85            &     0.84           & 0.81     \\
          $E_{\mathrm{max}}/E_{0}$                         &     4.4         &    4.8            &     5.1           & 5.0     \\
          $\mathrm{Q_{0}}$                  &     $2.91\times10^4$          &   $2.41\times10^4$             &     $1.91\times10^4$          & $1.42\times10^4$    \\
          Synchronous phase [deg.]                  &     \multicolumn{4}{c}{ -30    }      \\
       \hline
   \end{tabular}
   \label{tbl:dawcavity}
\end{table}%
The beam dynamics is designed using PARMILA~\cite{bib:parmila} and TRACE3D~\cite{bib:t3d} based on the designed cavity performance.
For ease of fabrication, a constant cell length \revc{($L=\beta_{s}\lambda/2$)} in a tank with multiple cells is designed. 
The design velocity $\beta_{s}$ and number of cells for each tank are determined based on the design of the beam dynamics. 
The inter-tank spacing is set to 4.5$\beta\lambda$ considering the feasibility of a magnet installation. 
The average acceleration field is determined to be 5.6~MV/m according to the Kilpatrick limit~\cite{bib:Kil57} and $E_{\mathrm{max}}/E_{0}$; 
the bravery factor is set to 0.9 in maximum. 
The number of cells in a tank is determined by the limitation of the quadrupole strength owing to transverse instabilities. 
There may be instabilities or resonances when the zero current phase advance ($\sigma_{0}$) is greater than 90 degrees~\cite{bib:Reiser94}. 
Although the muon beam intensity is expected to be much smaller than that in the region of such instabilities, the number of cells is chosen to match this criterion. 
The maximum $\sigma_{0}$ is 83~degrees at the first tank when the number of cells is ten, which determines the number of cells for all tanks. 
Table~\ref{tbl:dawtanks} shows the basic parameters of each DAW tank. 
Because the shunt impedance of the DAW cell increases as a function of $\beta$, 
the power required for the tank decreases with a higher energy. 
The total power required is 4.5~MW. 
The {\it phase slippage} is the greatest in the first tank, ranging from $-14$ to $+10$ degrees.
\begin{table}[h!tbp]
   \centering
   \caption{Parameters of each DAW tank.}
   \begin{tabular}{lrrrrr}
       \hline
           tank & $N_{\mathrm{cells}}$ & $\beta_s$ & length [m] & Energy [MeV] & Power [MW]                                       \\
       \hline
       	1     & 10 & 0.29 & 0.34 & 5.6 & 0.39 \\
       	2     &$\uparrow$& 0.33 & 0.38 & 7.1 & 0.35 \\
       	3    &$\uparrow$ & 0.37 & 0.42 & 8.8 & 0.33 \\
       	4   &$\uparrow$  & 0.40 & 0.46 & 10.7 & 0.31 \\
       	5     &$\uparrow$& 0.43 & 0.50 & 12.7 & 0.30 \\
       	6     &$\uparrow$& 0.47 & 0.54 & 14.9 & 0.29 \\
       	7     &$\uparrow$& 0.50 & 0.57 & 17.2 & 0.29 \\
       	8     &$\uparrow$& 0.52 & 0.61 & 19.7 & 0.29 \\
       	9     &$\uparrow$& 0.55 & 0.64 & 22.3 & 0.28 \\
       	10   &$\uparrow$& 0.58 & 0.67 & 25.0 & 0.28 \\
       	11   &$\uparrow$& 0.60 & 0.69 & 27.9 & 0.28 \\
       	12     &$\uparrow$& 0.62 & 0.72 & 30.8 & 0.28 \\
       	13     &$\uparrow$& 0.64 & 0.74 & 33.8 & 0.28 \\
       	14     &$\uparrow$& 0.66 & 0.77 & 37.0 & 0.28 \\
       	15     &$\uparrow$& 0.68 & 0.79 & 40.2 & 0.28 \\
       \hline
   \end{tabular}
   \label{tbl:dawtanks}
\end{table}
\par
\begin{figure}[!h]
\centering\includegraphics[width=5in]{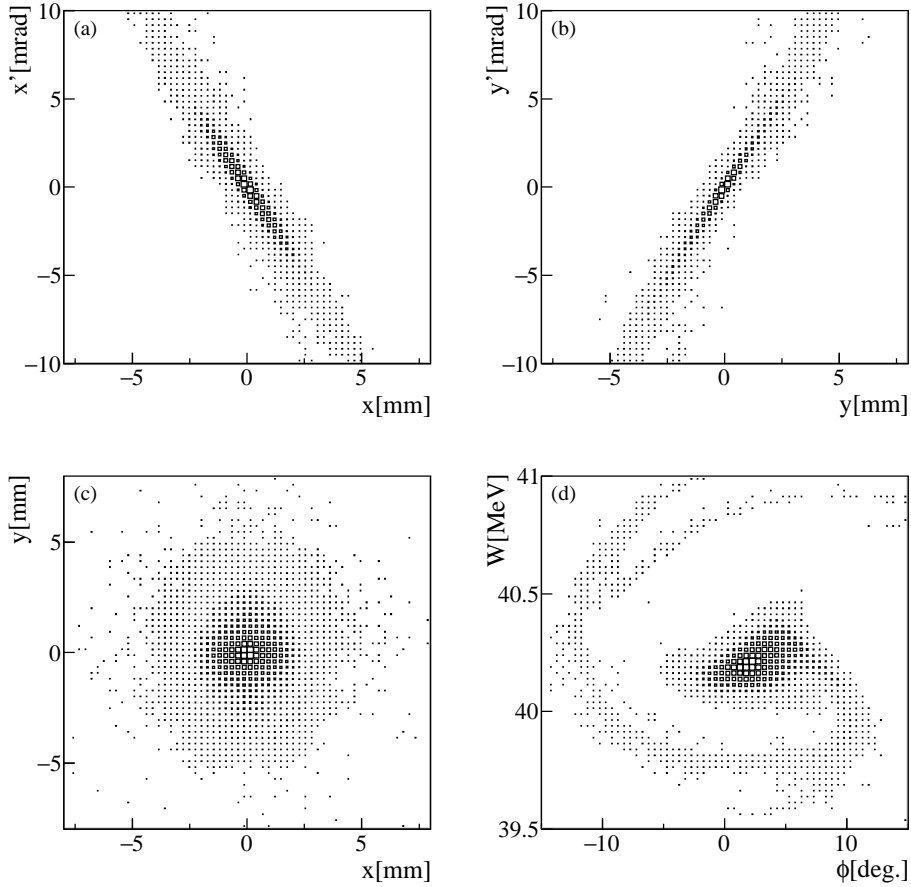}
\caption{Phase space distributions at the DAW-CCL exit: (a) Horizontal divergence angle x' 
vs. x, (b) vertical divergence angle y' vs. y, (c) y vs. x, and (d) $\Delta W$ vs. \revc{rf phase ($=$ time)}}
\label{fig:dawsim}
\end{figure}

The input distribution was given from the upstream IH-DTL simulation and used for the beam dynamics simulation after the calculation for the beam transport line to match the beam to the DAW-CCL. 
Figure~\ref{fig:dawsim} shows the phase space distributions at the DAW exit, and 
Table \ref{tbl:dawsim} summarizes the output beam parameters. 
The total length is 16.3~m with 15 modules, which corresponds to 138$\beta\lambda$, and the beam transit time \rev{$t_{\mathrm{trans.}}$} is calculated as $\frac{138\beta\lambda}{\beta c}=106.0$~ns. 
The survival probability is calculated as $\exp(t_{\mathrm{tran.}}/\tau_{\mu}\overline{\gamma})=96.1$\%, where $\overline{\gamma}=(\gamma_{\mathrm{in}}+\gamma_{\mathrm{out}})/2=1.212$. 
The emittance growth is estimated to be less than a few percent, and the output emittance is 0.32~$\pi$ and 0.21~$\pi$~ mm~mrad 
for the horizontal and vertical directions, respectively. 
The effect of errors on the emittance owing to a misalignment of the quadrupole magnet, power errors in the cavity, and other factors is evaluated to be less than 5\%~\cite{bib:takeuchi19}.
\begin{table}[h!tbp]
   \centering
   \caption{Output beam parameters of the DAW-CCL. \revc{In this table, $\alpha$ and $\beta$ are Twiss parameters. } }
   \begin{tabular}{ll}
       \hline
                                     & Output                                       \\
       \hline
           $\alpha_x$                                                      & 4.1                                \\
           $\beta_x$ [mm/mrad]                                     &  0.23            \\
           $\varepsilon_x$ [$\pi$~mm~mrad, rms, normalized ]   & 0.332 \\
           $\alpha_y$        & -6.9 \\
           $\beta_y$ [mm/mrad]         & 0.36                 \\
           $\varepsilon_y$ [$\pi$~mm~mrad, rms, normalized ]   & 0.201 \\
           $\alpha_z$        & -0.002                              \\
           $\beta_z$[deg/MeV]         & 24                        \\
           $\varepsilon_z$[$\pi$~MeV~deg, rms, normalized ]  & 0.108 \\
       \hline
           Transmission       & 99.80\%                                \\
           Transient time     & 106 ns                                \\
           Survival rate      & 96.1\%                               \\
           Transmission total & 95.9\%                                \\
       \hline
   \end{tabular}
   \label{tbl:dawsim}
\end{table}
\par
\begin{figure}[!h]
\centering\includegraphics[width=0.95\textwidth]{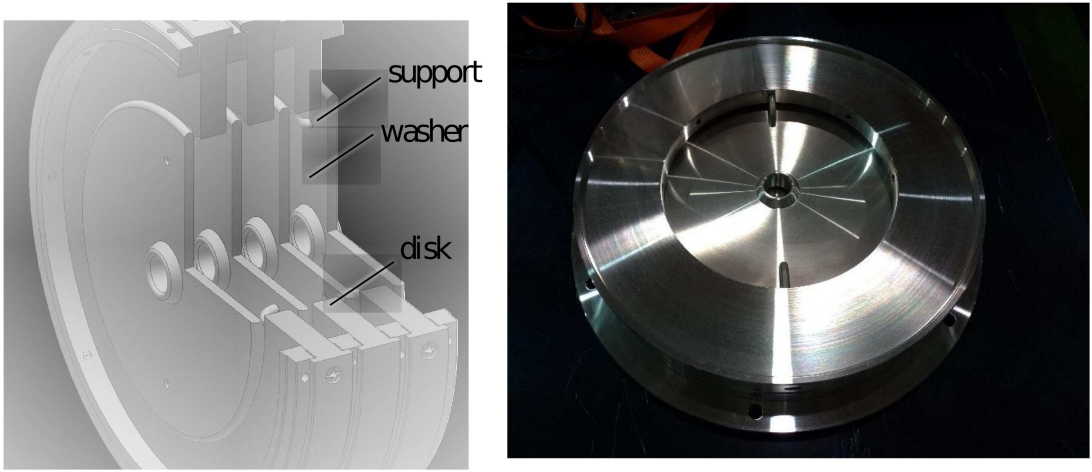}
\caption{Prototype of the DAW-CCL:  (Left) mechanical three-dimensional drawing, (Right) assembly of the prototype. 
\rev{The end plate with the antenna ports is attached for the low power measurements.}}
\label{fig:dawproto}
\end{figure}

A cold model of the first DAW cells was fabricated to confirm the design. 
\rev{The cold model is made of aluminum. }
Figure~\ref{fig:dawproto} shows a mechanical drawing (left) and photograph of the assembly (right). 
\rev{The resonant frequency was measured and consistent with the calculation using CST-MW Studio at 0.4\%. }
The on-axis electric field distribution was measured and the variation in the fields per cell was observed, which is considered to be due to the assembly process of the cavity. 
Based on the measurement results in the cold model, the actual DAW-CCL has been designed and the first tanks will be soon fabricated.
\subsection{DLS}
Finally, muons are accelerated to 212~MeV using the DLS. 
\par
\rev{DLS is classified as a radio-frequency linac for electron. 
After the first invention of the radio-frequency linac by Wider\"{o}e, Sloan and Lawrence~\cite{bib:Wid28, bib:Law31}, Hansen studied an electromagnetic field with radio-frequency resonator for accelerating electron~\cite{bib:Hansen38}. 
The acceleration of electron using a traveling wave accelerator was demonstrated in late 1940s~\cite{bib:Gin48}. }
Recent electron accelerators using room-temperature cavities have been based on the results of the Mark III linac~\cite{bib:Cho55} and SLAC. 
Unlike linear accelerators for electrons, which quickly reach the speed of light, muon linacs, which slowly approach the speed of light, require a gradual change in the length of the cell. 
\par
\begin{figure}[!h]
\centering\includegraphics[width=4in]{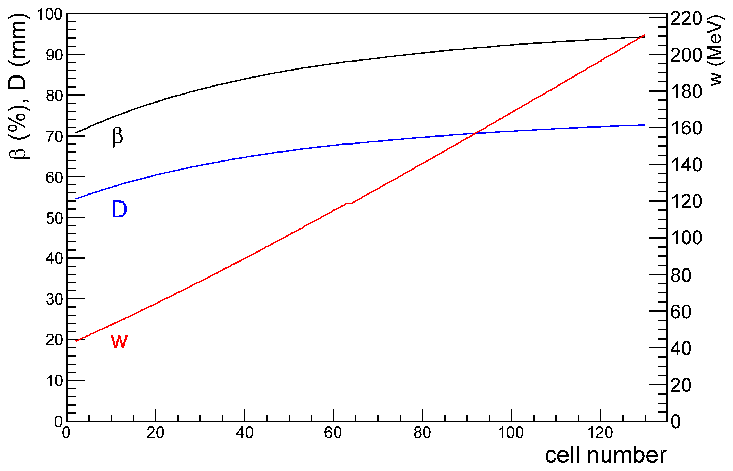}
\caption{Calculated cell parameters of the DLS section. $D$ is the cell length calculated by $\beta\lambda/3$ \rev{and $w$ is the kinetic energy of the muons.}}
\label{fig:dlsparam}
\end{figure}
\begin{figure}[!h]
\centering\includegraphics[width=5in]{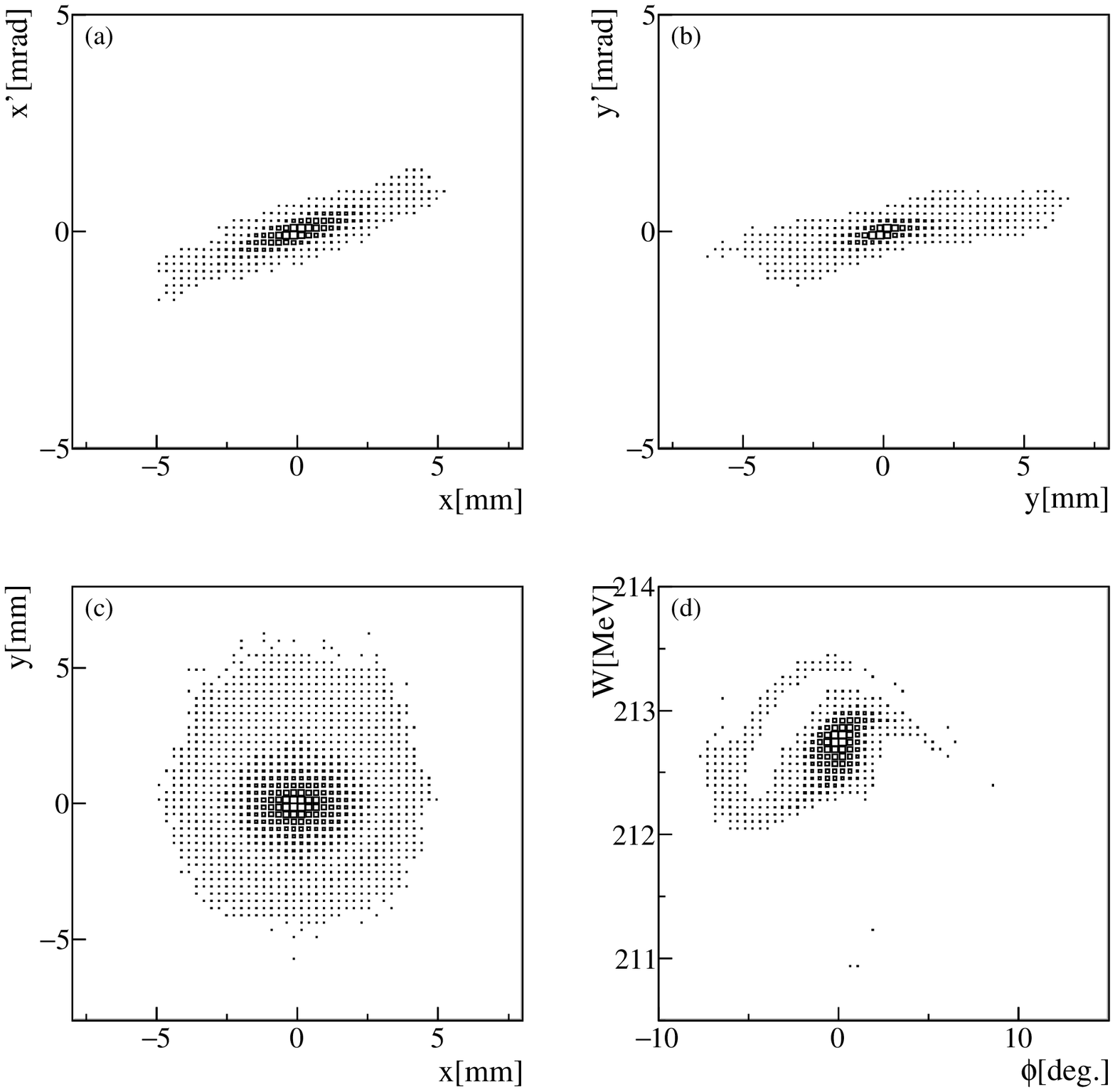}
\caption{Phase space distributions at the DLS exit: (a) Horizontal divergence angle x' 
vs. x, (b) vertical divergence angle y' vs. y, (c) y vs. x, and (d) $\Delta W$ vs. \revc{rf phase ($=$ time)}}
\label{fig:dlssim}
\end{figure}

The geometrical parameters of the DLS cell are designed using \rev{Poisson Superfish}. 
\rev{Poisson Superfish} generates the standing-wave-mode electric fields of each cell with open-open and short-short boundary conditions and 
the electric field of the traveling wave is represented by superposing these two fields with a phase difference of $\pi/2$. 
The L-band structure was adopted to make the acceptance sufficiently large for the input muon beam and 
the conventional $2\pi/3$ acceleration mode is adopted. 
The synchronous phase is set to $-10$ degrees to ensure sufficient longitudinal acceptance, and 
the average acceleration field ($E_0$) is assumed to be 20~MV/m. 
Figure~\ref{fig:dlsparam} shows the parameters of the DLS cells. 
In this study, a constant impedance design was adopted for simplicity. 
The calculated fields are implemented in the beam dynamics simulation conducted using the GPT. 
Figure~\ref{fig:dlssim} shows the phase space distributions at the DLS exit, and 
Table \ref{tbl:dlssim} summarizes the output beam parameters. 
Almost no emittance growth is observed. 
The transmission through the DLS section is \rev{99.9}\%, and the loss owing to the muon decay is estimated to be 1\%. 
\begin{table}[h!tbp]
   \centering
   \caption{Output beam parameters of the DLS. \revc{In this table, $\alpha$ and $\beta$ are Twiss parameters. } }
   \begin{tabular}{ll}
       \hline
                                     & Output                                       \\
       \hline
           $\alpha_x$                                                      & -1.9                                \\
           $\beta_x$ [mm/mrad]                                     &  0.86            \\
           $\varepsilon_x$ [$\pi$~mm~mrad, rms, normalized ]   & 0.331 \\
           $\alpha_y$        & -3.3 \\
           $\beta_y$ [mm/mrad]         & 1.5                 \\
           $\varepsilon_y$ [$\pi$~mm~mrad, rms, normalized ]   & 0.211 \\
           $\alpha_z$        & 0.07                              \\
           $\beta_z$[deg/MeV]         & 150                        \\
           $\varepsilon_z$[$\pi$~MeV~deg, rms, normalized ]  & 1.94 \\
       \hline
           Transmission       & 99.9\%                                \\
           Transient time     & 30 ns                                \\
           Survival rate      & 99\%                               \\
           Transmission total & 99\%                                \\
       \hline
   \end{tabular}
   \label{tbl:dlssim}
\end{table}
\subsection{Summary of the muon linac design}
\begin{figure}[!h]
\centering\includegraphics[width=0.95\textwidth]{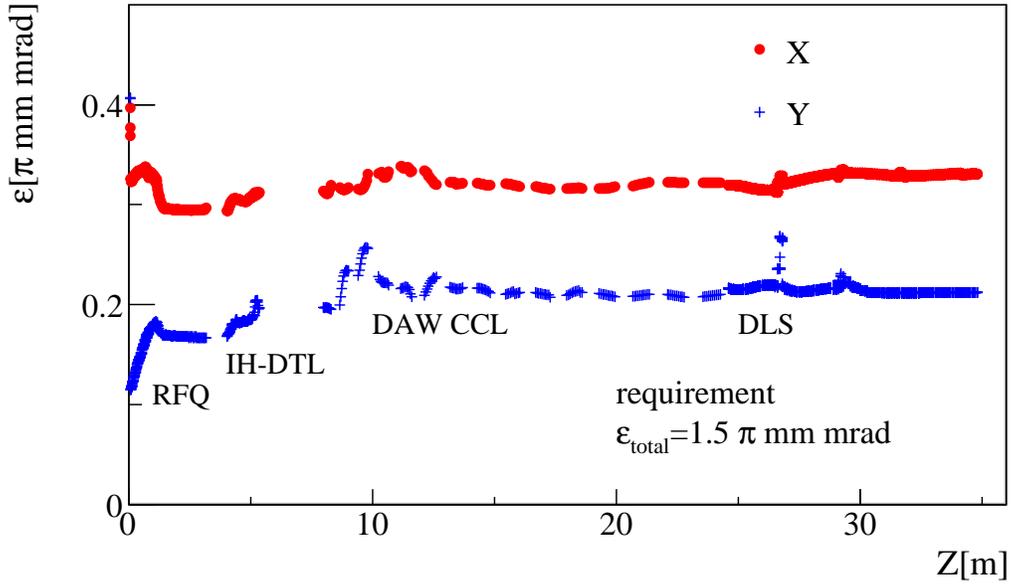}
\caption{Emittance evolution from the RFQ entrance to the linac exit. }
\label{fig:emitlinac}
\end{figure}

Figure~\ref{fig:emitlinac} shows the emittance evolution in the entire muon linac. 
There is no significant growth of the beam emittance and the output emittance is comparable to the \rev{required total emittance of 1.5$\pi$~mm~mrad}.  
Table~\ref{tbl:linacsum} summarizes the transmission, decay loss, and emittance at each section. 
The intensity of the low emittance muon beam at the linac exit is estimated to be \rev{$1.2\times10^4$ muons per pulse with a repetition of 25-Hz}. 
\rev{Table~\ref{tbl:allsum} shows the breakdown of the estimated transmission efficiency for all the experimental components and} 
the statistical uncertainty of the \amu~measurement is estimated to be 450~ppb for $2.2\times10^7$ s of data taking~\cite{bib:Abe20}. 
The statistical precision is comparable to that of the BNL experiment, and we are able to examine the \amu~ anomaly with a completely different scheme.
\par
\begin{table}[h!tbp]
   \centering
   \caption{Summary of the particle simulations through the muon linac: Transmission and decay loss in each section, and the emittance at the exit of each section. 
   \rev{The transmission in SOA includes that of the mesh electrodes. }
}
   \begin{tabular}{lccccc}
       \hline
                                                                                  & Soa  & RFQ  & IH-DTL & DAW-CCL & DLS                                       \\
       \hline
	Transmission \rev{[\%]} &                                                       \rev{56}      &  95  & 100 & 100  & 100 \\
	Decay loss \rev{[\%]}   &                                                       17      &  19  & 1     & 4     & 1\\
	$\varepsilon_x$ [$\pi$~mm~mrad, rms, normalized]&0.38   & 0.30& 0.32 & 0.32 & 0.33\\
	$\varepsilon_y$ [$\pi$~mm~mrad, rms, normalized]&0.11   & 0.17& 0.20 & 0.21 & 0.21\\
       \hline
   \end{tabular}
   \label{tbl:linacsum}
\end{table}
\begin{table}[h!tbp]
   \centering
   \caption{\rev{Breakdown of the estimated transmission efficiency. For the \amu\ measurement, $e^{+}$ energy window (12\%), detector acceptance (100\%), and reconstruction efficiency (12\%) should be considered. Details are described in~\cite{bib:Abe20}. }}
   \begin{tabular}{lcccccc}
       \hline
                                                                                  & H-line  &  \begin{tabular}{c}muonium \\emission\end{tabular}  & \begin{tabular}{c}laser \\ionization \end{tabular}& acceleration & injection & kicker                              \\
       \hline
	Transmission [\%] &                                                16        &  0.34             & 73                    & 40   & 84 & 93\\
       \hline
   \end{tabular}
   \label{tbl:allsum}
\end{table}
Because the systematic uncertainty is estimated to be less than 70~ppb and the measurement is statistically limited, 
further improvement of the beam intensity directly impacts the sensitivity of the physics. 
\rev{The goal in the next phase in the J-PARC \revr{E34} experiment is to measure \amu~with an accuracy of 100~ppb, comparable to the goal of the FNAL experiment. 
As shown in Table~\ref{tbl:allsum}, primary improvement should be in the muonium emission efficiency. 
The next one is the H-line and the acceleration efficiency. 
According to the results of previous muonium measurements~\cite{bib:Bak13, bib:Beer14, bib:Bea20}, 
the arrangement of the silica aerogel target is being investigated to enhance the overlap between the effective area of the target and the laser ionization region.
}
In the muon linac, the most effective way for the improvement is a reduction of the decay loss in the low-$\beta$ section. 
For this purpose, several efforts have been devoted: a new RFQ dedicated to the muons~\cite{bib:Kondo15}, an L-band RFQ combined with CC-DTL~\cite{bib:kondo20, bib:kondo18linac}, 
and a new scheme of acceleration for the low-energy muons~\cite{bib:otani20}. 
\rev{}
\par
Further, development is carried out for spin tracking~\cite{bib:yasuda20}. 
Because a spin precession in the muon storage ring starts with the initial spin state determined by the muon linac exit, 
the initial state should be understood to examine the potential systematic uncertainties of the physics measurement. 
For this purpose, a spin tracking simulation is being developed. 
In addition, a spin-rotator based on the Wien-filter type is designed to enable a spin-flipping analysis~\cite{bib:yasuda20jparc}, which is 
usually used in many types of spin experiments. 
\section{Demonstration of muon acceleration}\label{sec:muacc}
As described in Section~\ref{sec:e34}, a low emittance muon beam is 
realized through thermal muon production and acceleration using the linac dedicated to the muons. 
The production of thermal muons has been developed for over a quarter of a century~\cite{bib:Mills86, bib:Chu88, bib:Nagamine95} and has matured sufficiently and can be used in experiments. 
Muon acceleration, however, was an unproven technology.
Therefore, a muon acceleration should have been demonstrated prior to the construction of the actual linac. 
In addition, beam monitors are needed to diagnose the muon beam with an unprecedented beam size.
\par
To demonstrate muon acceleration prior to the construction of the experimental 
setup that included the thermal muon source system, a faster and simpler system for slow muons was needed. 
The scheme of muon cooling using a simple metal degrader and
acceleration using an RFQ proposed at Los Alamos National Laboratory~\cite{bib:Miyadera07} 
are suitable for this purpose. 
We basically followed this method, but to further reduce the emittance and separate accelerated muons from background 
high energy muons that pass through the RFQ without being accelerated, we decided to use negative muonium ions 
($\mu^{+} e^{-} e^{-}$; \mum) produced through an electron capture process~\cite{bib:Kua89}. 
Prior to the demonstration of muon acceleration by \mum, an experiment dedicated to a \mum~ measurement was carried out 
to evaluate the expected amount of accelerating \mum~ signal.
\par
This chapter describes the \mum~ measurement followed by a demonstration of muon acceleration along with a description of the development of the beam monitor.
\subsection{Development of \mum~source}
As mentioned above, the \mum~production process can be used to cool a muon beam down to 1~keV 
by simply irradiating positive muons ($\mu^+$'s) onto a thin metal film such as aluminum. 
However, since the first observation during the 1980s~\cite{bib:Kua89, bib:Har86}, there have been no measurements or no proof that a significant intensity 
can be obtained for an acceleration test. 
Therefore, we decided to conduct the \mum~measurement prior to the muon acceleration test. 
\par
\begin{figure}[!h]
\centering\includegraphics[width=0.85\textwidth]{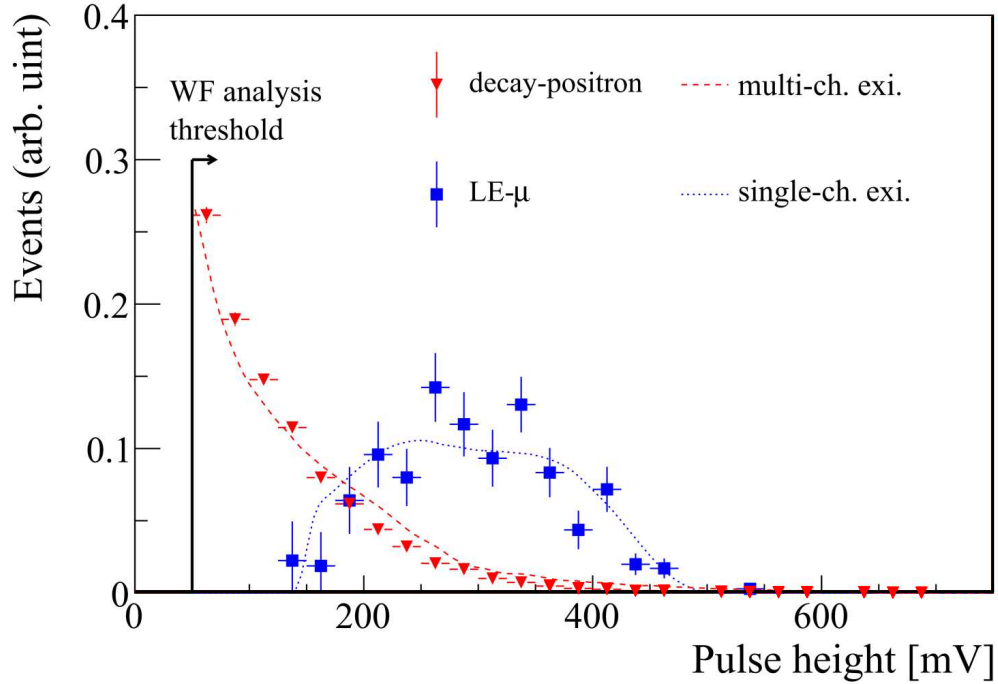}
\caption{Pulse height distribution for decay-positrons (red triangle) and LE-$\mu$ (blue box). 
Red dashed line and blue dotted line show the pulse height distribution with multi- and single-channel excitation models, respectively, as discussed in~\cite{bib:Otani19mcp}.
\rev{This figure is cited from~\cite{bib:Otani19mcp}. }}
\label{fig:phbgmu}
\end{figure}

 
In addition to the estimation of the accelerated muon signal, 
the method for identifying the background caused by positrons from muon decay (decay-positron) was also important. 
This is because the conversion efficiency from the primary $\mu^+$ to \mum~ is expected to be less than $10^{-4}$, which results in a large amount of decay-positrons derived from muon decays at the target. 
Therefore, the decay-positron background and low energy muon (LE-$\mu$) measurements were carried out~\cite{bib:Otani19mcp} 
using a micro-channel plate (MCP) detector  (Hamamatsu photonics, F1217-01~\cite{bib:hamamatsu}) in advance of the \mum~ measurements. 
In the decay-positron measurement, the decay-positrons from a muon beam target were identified by a series of triple scintillation detectors installed in front of the MCP detector. 
In the LE-$\mu$ measurement, the 7-keV $\mu^+$s are measured using the same setup as the \mum~measurement described below.
Figure~\ref{fig:phbgmu} shows the pulse height distribution
of the decay-positron (red triangle) and LE-$\mu$ (blue box). 
The difference between the decay-positrons and LE-$\mu$ in the pulse height distributions was
evident, which can be explained by the single- and multi- channel amplification model described in Section 4 of ~\cite{bib:Otani19mcp}. 
The results show that event selection using the pulse height can reject the decay-positron more than 75\% while maintaining the muon efficiency at 90\%. 
\par
\begin{figure}[!h]
\centering\includegraphics[width=5in]{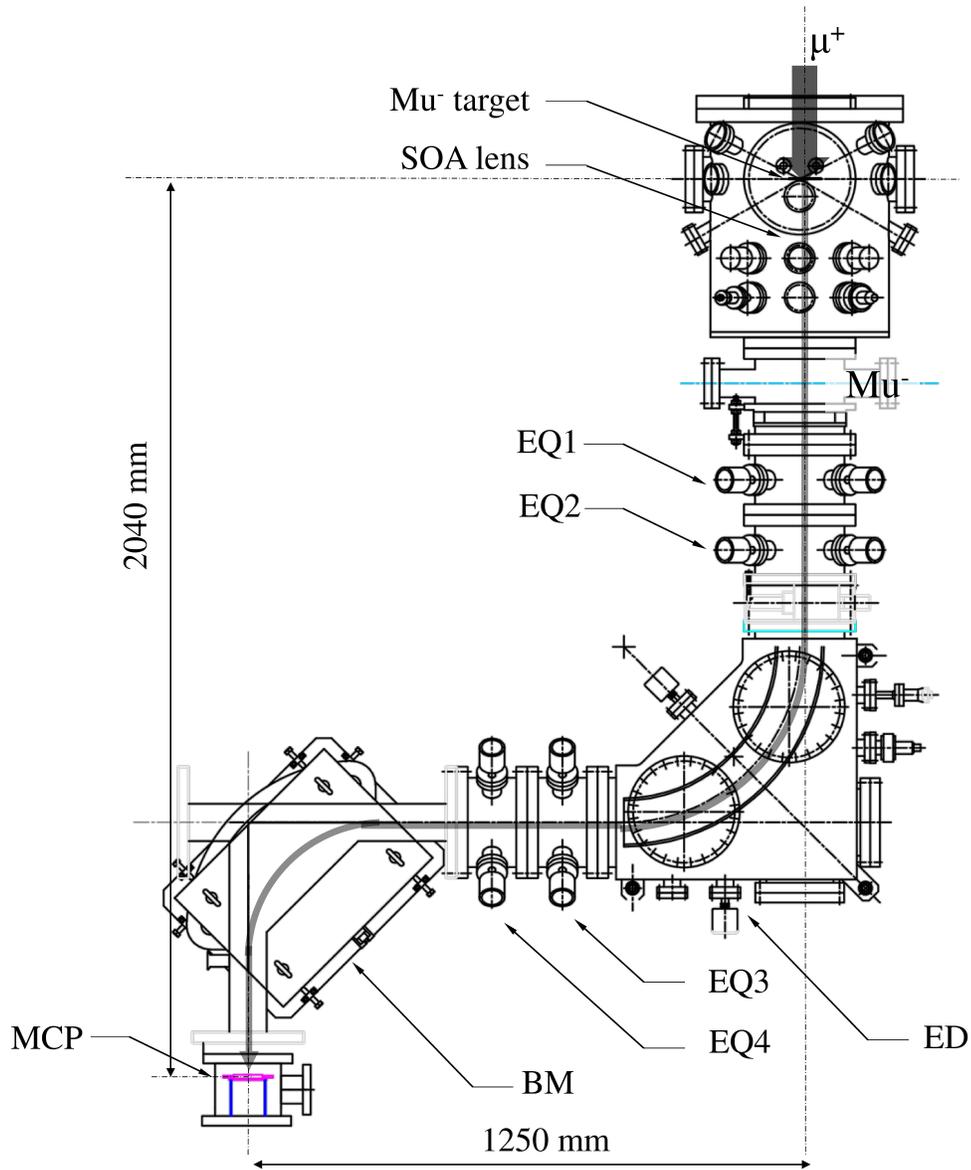}
\caption{Schematic of the typical experimental setup for the \mum~ measurements. 
\revr{This figure is cited from~\cite{bib:Otani19mum}.}}
\label{fig:mumsetup}
\end{figure}

The \mum~ measurements were conducted in a series of three experiments at the J-PARC MLF 
(2015A0324~\cite{bib:Otani19mcp}, 2016A0067~\cite{bib:rkita21}, and 2018B0007~\cite{bib:Otani19mum}). 
Figure~\ref{fig:mumsetup} shows a typical setup for the experiments. 
The $\mu^+$'s were injected into a \mum~production target after passing through a \rev{steel} window. 
The \mum~generated in the target was accelerated at up to 20~keV by the SOA electrostatic lens. 
Then, the \mum~was transported to the detector location through a series of electrostatic quadrupole (EQ1-4), an electrostatic deflector (ED), and a bending magnet (BM). 
The energy acceptance of the beamline is estimated to be 1.4\% by the GEANT4 simulation. 
The MCP detector was employed to measure the time of flight (TOF) from the \mum~ production target. 
The electrical signal from the MCP detector was amplified using a fast-filter amplifier (ORTEC 579) and 
digitized using CAEN V1720. 
The waveform within an interval of 10~$\mu$sec around each 25-Hz beam pulse was recorded for analysis. 
A pulse higher than the noise level was regraded as a signal pulse. 
The leading edge of the signal pulse was defined as the signal timing. 
The pulse height is defined by the maximum height within the signal window of 40~ns. 
The $\mu^+$ arrival time at the \mum\ production target was measured with a set of scintillating counters located at the side of the \mum\ production target. 
\par
In the 2015A0324 experiment, the LE-$\mu$ measurement was conducted to 
commission the experimental setup and compare the MCP signal to the decay-positron signal as described above. 
The aluminum (Al) thin foil is adopted as the target and LE-$\mu$ is extracted instead of \mum~by flipping the beamline polarity and transported to the MCP detector. 
The LE-$\mu$ signal was used to tune the applied voltage of the quadrupole and other experimental setups. 
In addition to the tuning using the LE-$\mu$, negative hydrogen ions generated by ultraviolet light~\cite{bib:nakazawa19} were used 
for offline commissioning of the experimental setup. 
\par
\begin{figure}[!h]
\centering\includegraphics[width=5in]{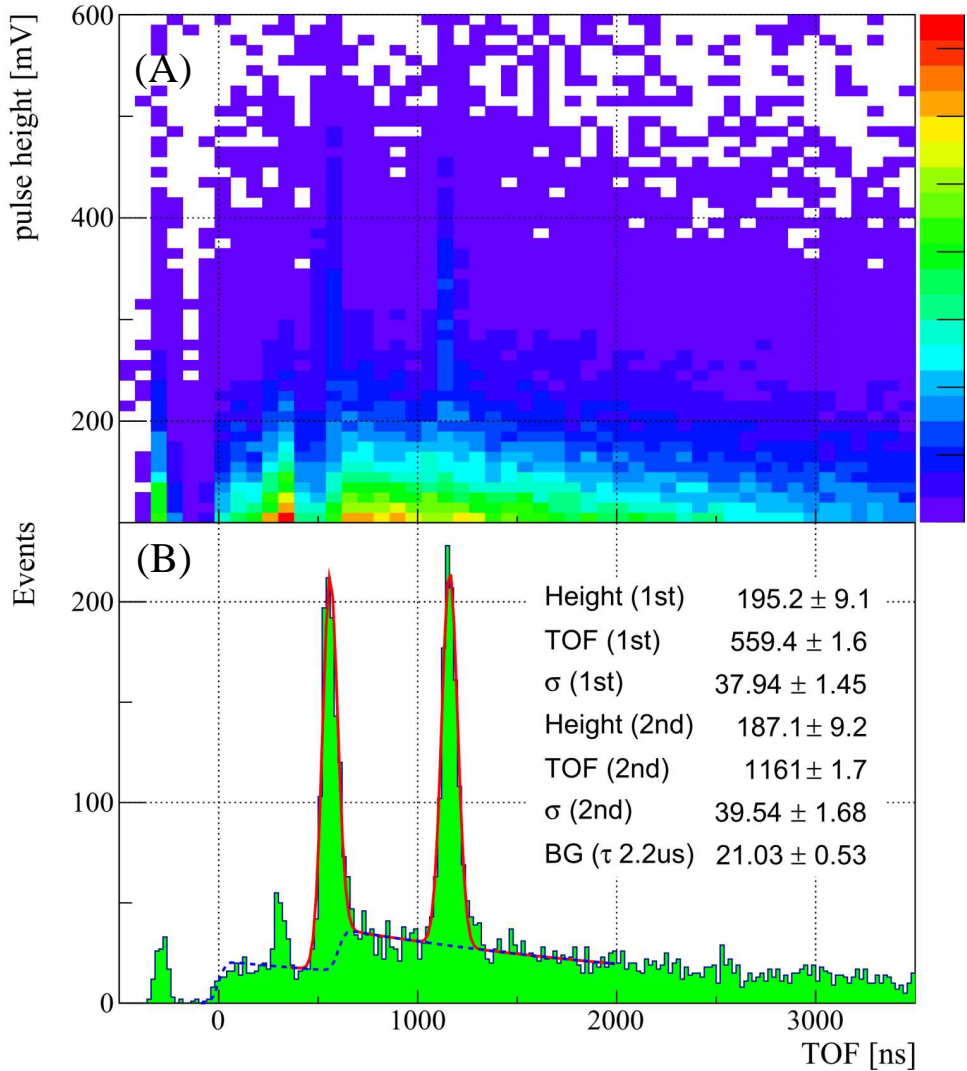}
\caption{(A) Pulse height vs. TOF for observed signal. (B) TOF distribution after pulse height selection. 
The solid red line and blue hatched line shows the fitting results assuming the \mum\ event and the decay-positron background. 
\revr{This figure is cited from~\cite{bib:Otani19mum}. }
}
\label{fig:tq}
\end{figure}

The \mum s were observed during the 2016A0067 experiment, and high-statistics \mum~data were obtained in the 2018B0007 experiment using three types of production targets: 
an Al foil, a C12A7 electride foil~\cite{bib:hosono03, bib:hosono17}, and a \rev{steel} foil. 
The electrical signals of the MCP detector were recorded and analyzed in a way similar to the 2015A0324 experiment. 
Figure~\ref{fig:tq} (A) shows a pulse height versus the TOF of the observed signal. 
The background events are considered to be decay-positrons from muons stopped in the experimental apparatus, 
based on the results of the analysis of time constants. 
The decay-positron events can be eliminated efficiently by the pulse height selection, as described above.
After pulse height selection, the TOF distribution is obtained as shown in Fig.~\ref{fig:tq} (B). 
The two peaks at approximately $-300$~ns and $300$~ns are due to prompt positrons carried through the surface muon beamline with the same momentum as $\mu^+$'s.
Because the prompt positron is faster than $\mu^+$, the prompt positron arrives earlier than $\mu^+$. 
The blue and red curves show the fitting results assuming the remaining decay-positron background (blue) and the \mum\ signals (red), respectively. 
The background is consistent with the exponential decay curve with the muon decay constant ($\tau_{\mu}=2.2$~$\mu$sec). 
The \mum~peak width is consistent with that of the primary $\mu^+$ beam. 
The time interval of the \mum~peaks is consistent with that of the primary proton beam pulses. 
The \mum~TOF is consistent within a few percentage points with the expectation estimated through the GEANT4 simulation where the initial energy of \mum\ is assumed to be 0.2~keV. 
\par
Based on the measured \mum~intensity during the 2016A0067 experiment, 
The expected intensity of the accelerated \mum~signal was estimated to be $6.8\pm1.2$ (stat.)$^{+0.4}_{-0.0}$ (syst.)/s~\cite{bib:rkita21}, \rev{where (stat.) represents the statistical error assigned by the measured \mum~intensity in the experiment and (syst.) represents the systematic error. }
The effect of a misalignment of the \mum~transport beamline is taken into account as a systematic error. 
From these results, it can be concluded that several hundred events of accelerated \mum~signals can be expected with a beam time of a few days, 
and that the background of the decay-positrons can be sufficiently suppressed by a pulse height discrimination, such that a sufficient amount of the accelerated \mum's can be observed with a sufficient signal-to-noise ratio. 
\subsection{First muon acceleration}
A demonstration of the muon acceleration was conducted in J-PARC MLF over a 6-day period starting on October 24, 2017 (2017A0263)~\cite{bib:Bae18}. 
\par
\begin{figure*}[!hbt]
\centering
\includegraphics[width = 0.95\textwidth]{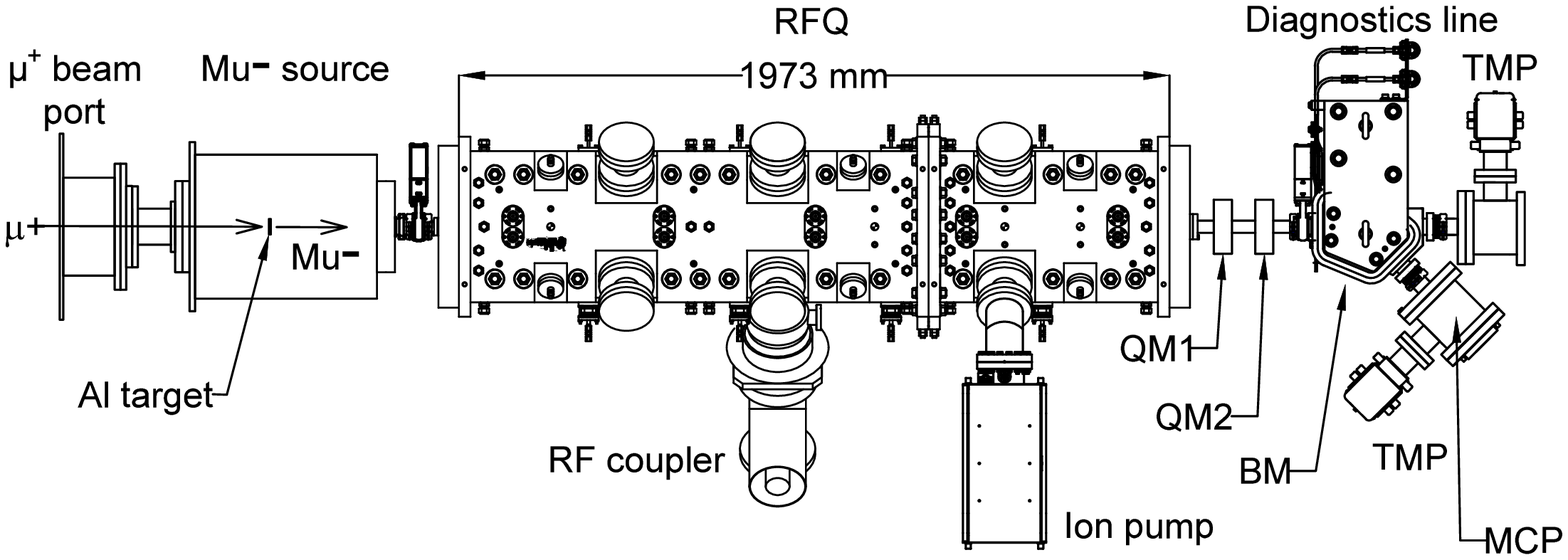}
\caption{Schematic drawing of the experimental setup for the demonstration of muon acceleration. 
\rev{In this figure, TMP is turbo molecular pump. }}
\label{fig:muaccsetup}
\end{figure*}

Figure~\ref{fig:muaccsetup} shows a schematic drawing of the experimental setup. 
The $\mu^+$'s were incident on the Al foil target used in the \mum~measurements. 
The $\mu^+$'s were decelerated through the target, and some of the 
$\mu^+$'s become Mu$^-$'s at the 
downstream surface of the Al target. 
Using the same Soa lens as used in the \mum measurements, the generated \mum's 
were accelerated to 5.6~keV and focused on the entrance of the RFQ. 
In this experiment, the prototype RFQ~\cite{bib:kondo06} was used. 
The length of this RFQ is equivalent to two-thirds of the RFQ in the muon linac, and 
this RFQ was originally designed using KEKRFQ~\cite{bib:Ueno90} to accelerate negative hydrogen ions at up to 810~keV. 
To use this RFQ for a muon acceleration, the intervane voltage should
be normalized to the muon mass, and the input velocity $\beta$ should
be the same as that of H$^-$. 
Table~\ref{tbl:xrfqpqr} summarizes the parameters of the prototype RFQ. 
The \mum's are accelerated to 89~keV and then transported to the MCP detector through the diagnostic beamline 
consisting of two quadrupole magnets (QM1 and QM2) and a bending magnet (BM). 
Because the expected number of \mum~ signals is only a few hundred, and it is impossible to obtain the correct field setting using the \mum~ signals themselves, 
the diagnostic beamline was commissioned using negative hydrogen ions generated by ultraviolet light~\cite{bib:nakazawa19}, and the field setting of the magnets was verified prior to the experiment. 
The electrical signals of the MCP detector are recorded in a similar way as the experiments of the \mum~measurements.
%
\par
\begin{table}[h!tbp]
   \centering
   \caption{Parameters of the prototype RFQ for the H$^{-}$ and muon acceleration.}
   \label{}
   \begin{tabular}{lll}
       \hline
                                                   & H$^-$ & muon             \\
       \hline
           Frequency (MHz) & \multicolumn{2}{c}{324}    \\
           Number of cells                & \multicolumn{2}{c}{297}            \\
           Length (m)                       & \multicolumn{2}{c}{1.97}          \\
           Intervane voltage (kV)       & 81    & 9.1               \\
           Power (kW)                       & 180    &  2.3              \\
           Injection energy (keV)        & 50  & 5.6                \\
           Extraction energy (keV)    & 810    & 89 \\
       \hline
   \end{tabular}
   \label{tbl:xrfqpqr}
\end{table}
At the start of the experiment, the polarity of the beamline was set to transport positively charged particles. 
In this positive-charge configuration, muons that passed through the target and were decelerated to 89 keV, which corresponds to the acceleration energy, were measured. 
The observed TOF is consistent with calculations based on the distance between the target and the MCP detector and the velocity of the 89-keV muon. 
This confirms the beam diagnostic system. 
\par
\begin{figure}[hbt]
\centering
\includegraphics[width=0.9\textwidth]{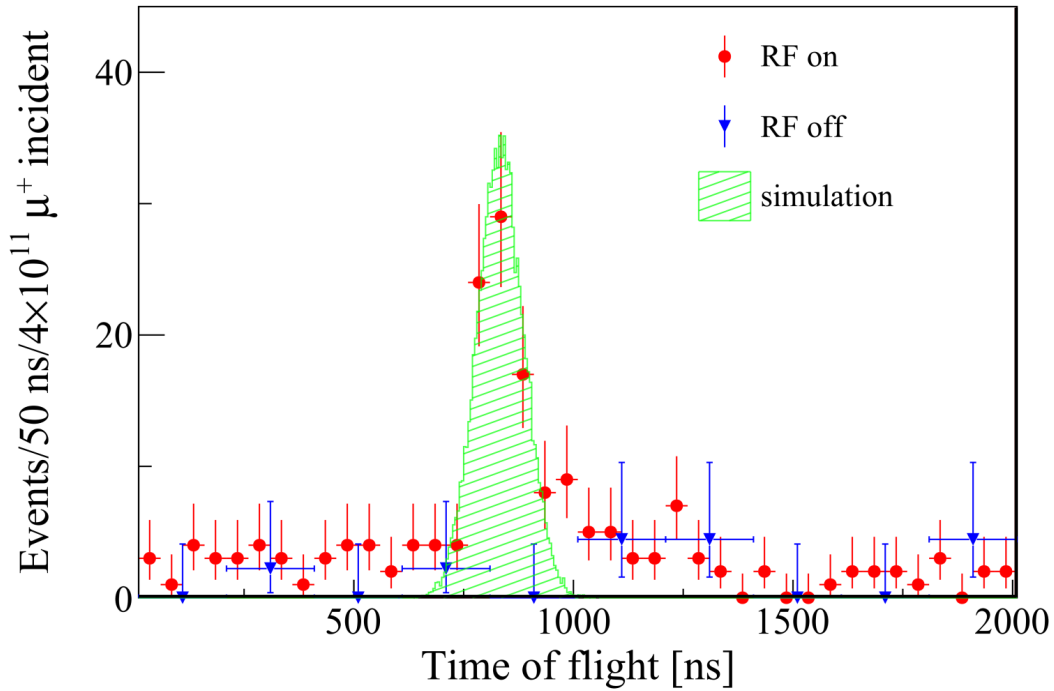}
\caption{TOF spectra of the negative-charge configuration with RF on and off.
  The clear peak of the RF on the spectrum at 830~ns corresponds to
  the accelerated Mu$^-$'s. The error bars are statistical.
  A simulated TOF spectrum of the accelerated Mu$^-$'s is also plotted. 
\rev{This figure is cited from~\cite{bib:Bae18}. } } 
\label{fig:tof}
\end{figure}

After measurement of 89-keV $\mu^+$'s, the polarities of the magnets were flipped to a negative-charge configuration. 
Figure~\ref{fig:tof} shows the TOF spectrum with and without
the RFQ operation after the pulse-height cut was applied.
With the RFQ operation, a clear peak was observed at $830\pm11$~ns. 
The time required to reach the RFQ entrance from the target while being accelerated by the Soa lens was estimated as 307~ns using the GEANT4 simulation. 
The number of cells of the prototype RFQ is 297, and thus it takes
$\frac{297}{2\times324\times10^6} = 458$~ns to fully accelerate the
particles through the 324~MHz field. 
The length of the diagnostic beamline is 0.91~m, and thus the transit time of the 89-keV Mu$^-$ is 72~ns. 
The total flight time of the accelerated Mu$^-$ from the target to the MCP detector was calculated as
$307 + 458 + 72 = 837$~ns, which is consistent with the measurement. 
The hatched histogram in Fig.~\ref{fig:tof} represents the simulated TOF spectrum of the accelerated Mu$^-$. 
The number of simulation events was normalized to the number of incident muons of this data set. 
The 46~ns rms width of the TOF spectrum is consistent with that
from the timing distribution of the primary $\mu^+$ at the Al target.
\par
From these experimental results, it is concluded that the observed
TOF peak is due to the Mu$^-$'s accelerated by the RFQ to 89~keV. 
The event rate was estimated as $(5\pm1) \times 10^{-4}$/s by subtracting the decay-positron events
estimated from the timing region outside the signal range. 
This is consistent with the expectation based on the \mum~ measurements. 
The intensity of the accelerated \mum~ is limited by the low conversion efficiency of $\mu^+$ to \mum. 
\par
\subsection{Development of beam monitors}\label{sec:monitor}
After the first demonstration of the muon acceleration described above, 
the beam monitors for the accelerated muons were tested using the accelerated \mum's. 
To diagnose the beam in both the transverse and longitudinal directions, 
a beam profile monitor (BPM) and a bunch width monitor (BWM) are necessary. 
\par
The BPM has been developed based on a MCP assembly (Hamamatsu F2225-21P) and a charge-coupled device (CCD) camera. 
The avalanche electrons generated by the MCP are injected into the phosphor screen, from which the light output is focused onto the CCD camera. 
The spatial resolution was estimated as 0.3~mm using an ultraviolet light and surface muons~\cite{bib:kim18}. 
During the 2016B0214 experiment, the detector in the setup for the \mum~measurement shown in Fig.~\ref{fig:mumsetup} was replaced with the BPM 
and the performance against LE-$\mu$ was demonstrated~\cite{bib:rkita17}. 
During the 2017B0006 experiment, the detector in the setup shown in Fig.~\ref{fig:muaccsetup} is replaced with the BPM 
and the beam profiles of the accelerated \mum's were measured. 
The Kolmogorov–Smirnov test was performed and the measurement and expectation were consistent within a statistical error (\rev{P-value is 21\% and 38\%} for the horizontal and vertical direction, respectively)~\cite{bib:Otani18muaccbpm}. 
\par
The BWM employed a MCP assembly (Hamamatsu F1217-11G) containing two stages of chevron-type MCPs. 
The signal-processing electronics of constant-fraction discriminators~\cite{bib:Les75}, 
which were adopted from the technology of the other experiment~\cite{bib:Inami14}, measured the timing of the MCP signal. 
It detects single muon with a high temporal resolution, accumulates muon events, and reconstructs the beam bunch 
by taking the difference from the RF reference time of the muon linac. 
A test bench was developed to evaluate the temporal resolution of the BWM using a picosecond pulse laser (Hamamatsu PLP10-040)~\cite{bib:yotsuzuka19}. 
Recent results have evaluated the resolution of the BWM to be 40 ps, which corresponds to 1\% of the acceleration frequency of 324 MHz and satisfies the requirement. 
During the \rev{2018A0222} experiment, the detector in the setup shown in Fig.~\ref{fig:muaccsetup} is replaced with the BWM to measure the bunch width of the accelerated \mum's. 
In addition, a buncher cavity developed as a prototype for the muon linac~\cite{bib:Otani19buncher} was installed between QM2 and BM to focus the beam in the longitudinal direction. 
The bunch width was successfully measured to be $0.54\pm0.11$~ns, which is consistent with the simulation~\cite{bib:sue20}. 
\par
From these measurements, the beam monitors are ready to diagnose the accelerated muon beam, particular for low energy part. 
Because these monitors are based on an MCP and the detection efficiency of an MCP is expected to decrease as the muon energy increases, 
to conduct sufficient beam commissioning, it is necessary to develop a beam monitor for the high-energy part. 
\section{Summary and outlook}\label{sec:sum}
\rev{
The design of the muon linac for the J-PARC E34 experiment and the first muon acceleration were presented. 
The experiment for demonstrating the acceleration of the thermal muons is scheduled in 2022 
at a new muon experiment area (S2) in J-PARC MLF. 
The experiment aims to study the laser ionization of the muonium via the 1S-2S excitation with 244-nm laser~\cite{bib:Ced21} and 
to accelerate the thermal muons using the RFQ. 
The muon beam emittance before and after the acceleration will be measured using 
the beam monitors described in \revr{Section}~\ref{sec:monitor}. 
After the experiment at S2, the muon acceleration using the RFQ and IH-DTL at the H-line 
will be conducted. 
}
\par
We are entering a new era in which the accelerated muons are available. 
One can naturally imagine an imaging technique with accelerated muons, allowing for a better resolution with less time than cosmic-ray muons. 
It is natural to think of a collider using an accelerated muon beam. 
A type of collider in which muons collide with other particles has been actively discussed~\cite{bib:Men20, bib:Kin21}. 
If a cooling method of negative muons with an efficiency comparable to that of the muonium laser ionization is realized, 
a muon collider will be promised. 
\par
The J-PARC E34 experiment will soon be realized as a flagship to new horizon for a better understanding of nature. 
\section*{Acknowledgment}
We express our appreciation to the many manufacturing companies
involved in this project, especially to Toshiba Co., who fabricated
the \rev{prototype} RFQ, \rev{Toyama Co.,Ltd., who fabricated the actual RFQ}, and 
\rev{TIME Co., Ltd.}, who fabricated the 
buncher cavity and the prototype IH-DTL. 
This work is supported by JSPS KAKENHI (Grant Numbers 
, 25800164
, 15H03666
, 15H05742
, 16H03987
, 16J07784
, 18H03707
, 18J22129
, 19J21763
, 20J21440
, 20H05625
, 21K18630
, 21H05088
, 22H00141
) 
, JST FOREST Program (Grant Number JPMJFR212O) 
and the natural science grant of the Mitsubishi Foundation. 
This work is also supported by
the Korean National Research Foundation grants
NRF-2015H1A2A1030275,
NRF-2015K2A2A4000092,
and NRF-2017R1A2B3007018;
the Russian Foundation for Basic Research
grant RFBR 17-52-50064; 
and the Russian Science Foundation grant
RNF 17-12-01036.
This experiment at the Materials and Life Science Experimental Facility
of the J-PARC was performed under user programs (Proposal No. 2015A0324, 2016A0067, 2016B0214, 2017A0263, 2017B0006, 2018A0222, and 2018B0007).
\let\doi\relax


\begin{thebibliography}{999}
\bibitem{bib:Bae18}
S. Bae et al., Phys. Rev. AB 21, 050101 (2018).
\bibitem{bib:And37}
S.H. Neddermeyer, C.D. Anderson, Phys. Rev. 51, 884 (1937).
\bibitem{bib:Mic75}
E. G. Michaelis,  IEEE Trans. Nucl. Sci.Vol. NS-23,  No. 3, 1385 (1975).
\bibitem{bib:Man19}
M. Boscolo, J. P. Delahaye and M. Palmer, Rev. Acc. Sci. Tech., Vol 10, No. 01, 189-214 (2019)
\bibitem{bib:Bos18}
M. Boscolo et al., Phys. Rev. AB 21, 061005 (2018).
\bibitem{bib:tmm}
\url{https://slowmuon.kek.jp/index_e.html}
\bibitem{bib:Abe20}
M. Abe et al., Prog. Theor. Exp. Phys. 2019, 053C02 (2019). 
\bibitem{bib:Cha62}
G. Charpak, F. J. M. Farley, R. L. Garwin, T. Muller, J. C.Sens, and A. Zichichi, Phys. Lett. 1, 16 (1962).
\bibitem{bib:Bai72}
J. Bailey et al., Nuovo Cimento A 9, 369 (1972).
\bibitem{bib:Bai79}
J. Baileyet al., Nucl. Phys. B 150, 1 (1979). 
\bibitem{bib:Ben06}
Bennett G. W. et al.  (Muon $g-2$ Collaboration), Phys. Rev. D 73, 072003 (2006).
\bibitem{bib:Aoyama20}
T. Aoyama et al., Phys. Rept. 887, 1   (2020). 
\bibitem{bib:Gra15}
J. Grangeet al. (Muon $g-2$ Collaboration), arXiv:physics.ins-det/1501.06858
\bibitem{bib:abi21}
B.~Abi et al. (Muon $g-2$ Collaboration), Phys. Rev. Lett. 126, 141801 (2021). 

\bibitem{bib:Str10}
D. Stratakis et al., Phys. Rev. AB 22, 011001 (2019). 
\bibitem{bib:Kawamura18}
N. Kawamura et al., Prog. Theor. Exp. Phys. 2018, 113G01 (2018).
\bibitem{bib:Bak13}
Bakule P. et al. , Prog. Theor. Exp. Phys. 2013, 103C01 (2013).
\bibitem{bib:Beer14}
G.A. Beer et al., Prog. Theor. Exp. Phys. 2014, 091C01 (2014).
\bibitem{bib:Bea20}
J Beare et al., Prog. Theor. Exp. Phys. 2020, 123C01 (2020).
\bibitem{bib:Iinuma16}
H. Iinuma et al.,  Nucl. Instrum. Meth. A 832, 51 (2016).
\bibitem{bib:Abe18}
M. Abe et al., Nucl. Instrum. Meth. A 890, 51 (2018).
\bibitem{bib:Aoyagi20}
T. Aoyagi et al., JINST 15, P04027 (2020).
\bibitem{bib:Kishishita20}
T. Kishishita et al., IEEE TNS, 67, 2089 (2020)
\bibitem{bib:Kondo13}
Y. Kondo et al., Phys. Rev. AB 16, 040102 (2013).
\bibitem{bib:Kubosaki11}
M. Kubosaki et al.,, Proceedings of PASJ 2011, 1366 (2011). 

\bibitem{bib:Kondo15}
Y. Kondo et al., Proceedings of IPAC 2015, 3801 (2015).
\bibitem{bib:Otani16}
M. Otani et al., Phys. Rev. AB 19, 040101 (2016). 
\bibitem{bib:Otani19daw}
M. Otani et al., J. Phys. :Conf. Ser. 1350, 012097 (2019).
\bibitem{bib:Kondo17}
Y. Kondo et al., J. Phys. :Conf. Ser. 875, 012054 (2017).

\bibitem{bib:Can86}
K. F. Canter, P. H. Lippel, W. S. Crane,  and A. P. Mills Jr., in ``Positron studies of solids, surfaces and atoms'' (World Scientiﬁc, Singapore, 1986) p. 199.
\bibitem{bib:Otani18_surfmu}
M. Otani et al., J. Phys. :Conf. Ser. 1067, 052018 (2018).
\bibitem{bib:g4bl}
G4beamline, \url{http://public.muonsinc.com/Projects/G4beamline.aspx}
\bibitem{bib:Stra10}
P. Strasser et al., Journal of Physics: Conference Series 225, 012050 (2010).
\bibitem{bib:g4}
GEANT4, \url{http://geant4.cern.ch/}
\bibitem{bib:opera}
OPERA3D, Vector Fields Limited, Oxford, England., \url{https://operafea.com/}

\bibitem{bib:parmteqm}
K. R. Crandall et al., LA-UR-96-1836 (2005).
\bibitem{bib:Kap70_1}
I.M.~Kapchinskiy and V.A.~Tepliakov, Prib. Tekh. Eksp 2, 19-22 (1970). 
\bibitem{bib:Kap70_2}
I.M.~Kapchinskiy and V.A.~Tepliakov, Prib. Tekh. Eksp 4, 17-19 (1970). 
\bibitem{bib:Sto81}
J.E.~Stovall, K.R.~Crandall, R.W.~Hamm, IEEE Trans. Nucl. Sci. NS-28  p.1508 (1981)
\bibitem{bib:Otani15_pasj}
M. Otani et al., Proceedings of PASJ 2015, 56 (2015)
\bibitem{bib:Mor49}
H. Morinaga, Phys. Soc. Meeting, Osaka (1949).
\bibitem{bib:Ble56}
J.P. Blewett, Proceedings of Symposium du CERN sur les Accelerateurs de Haute Enegie et la Physique des Mesons $\pi$ v.1, 162 (1956). 
\bibitem{bib:Zei62}
P.M. Zeidlitz and V.A. Yamnitskii, J. Nucl. Energy, Part C4, 121 (1962).
\bibitem{bib:Pot69}
J. Pottier, IEEE Trans. Nucl. Sci. 16/3, 377 (1969). 
\bibitem{bib:Nol79}
E. Nolte et al., Nucl. Instr. and Meth. 158, 311 (1979).

\bibitem{bib:Rat05}
U. Ratzinger, CERN Yellow Report, 2005-003, 351 (2005).
\bibitem{bib:Good53}
Good, M.L., Phys. Rev. 92, 538 (1953).
\bibitem{bib:Min99}
S. Minaev and U. Ratzinger, Proceedings of PAC Conf 1999., 3555 (1999).
\bibitem{bib:Iwata06}
Y. Iwata et al., Nucl. Instr. Meth. A569, 685 (2006).
\bibitem{bib:jameson14}
R. A. Jameson, arXiv:physics.acc-ph/1404.5176
\bibitem{bib:cst}
CST Studio Suite, Computer Simulation Technology (CST). \url{https://www.cst.com/products/CSTMWS}
\bibitem{bib:Otani16pasj}
M. Otani et al., Proceedings of the PASJ 2016, 858-862 (2016). 
\bibitem{bib:gpt}
General Particle Tracer, Pulsar Physics. \url{http://www.pulsar.nl/gpt/}

\bibitem{bib:Nakazawa19}
Y. Nakazawa et al., J. Phys. :Conf. Ser. 1350, 012054 (2019).
\bibitem{bib:bpm}
Peter A. Mcintosh, Proceedings of 4th European Particle Accelerator Conference, 1283 (1994). 
\bibitem{bib:and72}
V. G. Andrev et al.,Proceedings of Linear Acel. Conf 1972, 114 (1972)
\bibitem{bib:Mur72}
B.P.~Murin et al.,  Proceedings of Proton Linac Conf 1972, LA-5115, p.387 (1972).
\bibitem{bib:and76}
V.G.Andreev et al., Proceedings of Linac Conf. 1976, AECL-5677 p.269 (1976). 
\bibitem{bib:Esi88}
S. K. Esin et al., Proceedings of Linear Accel. Conf 1988. 657 (1988) 
\bibitem{bib:Ina86}
S. Inagaki, Nucl. Instrm. and Meth. in Phys. Res. A 251, 417-436 (1986).
\bibitem{bib:Iwa94}
Y. Iwashita, Nucl. Instrm. and Meth. in Phys. Res. A 348, 15-33 (1994).
\bibitem{bib:sf}
J.H. Billen and L.M. Young, Los Alamos Report, LA-UR-96-1834 (1996).
\bibitem{bib:ao00}
H. Ao et al., Jpn. J. Appl. Phys. 39, 651 (2000).
\bibitem{bib:parmila}
\url{http://www.laacg.lanl.gov}
\bibitem{bib:t3d}
K.R. Crandall and D.P Rustoi, Los Alamos Report, LA-UR-97-886 (1997).
\bibitem{bib:Kil57}
W.D. Kilpatrick, Rev. Sci. Instr. 28, 824 (1957). 
\bibitem{bib:Reiser94}
Reiser M ``Theory and Design of Charged Particle Beams''1994 JOHN WILEY \& SONS, INC.
\bibitem{bib:takeuchi19}
Y. Takeuchi et al., to be published in proceedings of J-PARC symposium. 

\bibitem{bib:Wid28}
R. Wider\"{o}e, Arch. Electrotech., 21, 387 (1928).
\bibitem{bib:Law31}
D.H.~Sloan and E.O.~Lawrence, Phys. Rev. 38, 2021-2032 (1931). 
\bibitem{bib:Hansen38}
H. Hansen, J. App. Phys. 9, 654 (1938)
\bibitem{bib:Gin48}
E.L.~Ginzton, W.W.~Hansen and W.R. Kennedy, Rev. Sci. Instrum, 19, 89 (1948)
\bibitem{bib:Cho55}
M. Chodorow et al., Rev. Sci. Instrum. 26, 134 (1955). 



\bibitem{bib:kondo20}
Y. Kondo, M. Otani, Proceedings of PASJ2020, 218 (2020) (in Japanese). 
\bibitem{bib:kondo18linac}
Y. Kondo, T. Morishita, J. Tamura, M. Otani, Proceedings of LINAC 2018, 794 (2018).
\bibitem{bib:otani20} 
M. Otani, Y. Kondo, Proceedings of PASJ2020, 202 (2020) (in Japanese). 

\bibitem{bib:yasuda20}
H. Yasuda et al., Proceedings of PASJ2020, 173 (2020) (in Japanese). 


\bibitem{bib:yasuda20jparc}
H. Yauda et al., to be published in proceedings of J-PARC symposium. 

\bibitem{bib:Mills86}
A. P. Mills, Jr., J. Imazato, S. Saitoh, A. Uedono, Y. Kawashima, and K. Nagamine
Phys. Rev. Lett. 56, 1463 (1986). 
\bibitem{bib:Chu88}
Steven Chu, A. P. Mills, Jr., A. G. Yodh, K. Nagamine, Y. Miyake, and T. Kuga
Phys. Rev. Lett. 60, 101 (1988). 
\bibitem{bib:Nagamine95}
K. Nagamine et al., Phys. Rev. Lett. 74, 4811 (1995)

\bibitem{bib:Miyadera07}
H. Miyadera et al., in proceedings of PAC 2007, 3032 (2007). 
\bibitem{bib:Kua89}
Y. Kuang et al., Phys. Rev. A39, 6109 (1989)

\bibitem{bib:Har86}
D.R. Harshman, et al., Phys. Rev. Lett. 56, 2850 (1986). 


\bibitem{bib:Otani19mcp}
M.~Otani et al., Nucl. Inst. Meth. Phys. Res. Sec. 946, 162693 (2019). 
\bibitem{bib:hamamatsu}
\url{http://www.hamamatsu.com/}

\bibitem{bib:rkita21}
R. Kitamura et al., to be published in Phys. Rev. AB.


\bibitem{bib:Otani19mum}
M. Otani et al., J. Phys. :Conf. Ser. 1350, 012067 (2019).


\bibitem{bib:nakazawa19}
Y. Nakazawa et al., Nucl. Inst. Meth. Phys. Res. Sec. 937, 164 (2019). 


\bibitem{bib:hosono03}
H. Hosono et al., Science, 301, 626 (2003). 

\bibitem{bib:hosono17}
H. Hosono et al., Proc. Natl. Acad. Sci. USA 114(2), 233 (2017). 

\bibitem{bib:kondo06}
Y. Kondo, K. Hasegawa,  and A. Ueno, Proceedings of LINAC 2006, 749 (2006).

\bibitem{bib:Ueno90}
A. Ueno and Y. Yamazaki, Proceedings of Linac Conf. 1990, 329 (1990).


\bibitem{bib:kim18}
B. Kim et al., Nucl. Inst. Meth. Phys. Res. Sec. A. 899, 22 (2018). 

\bibitem{bib:rkita17}
R Kitamura et al., J. Phys. :Conf. Ser. 875, 012055 (2017).



\bibitem{bib:Otani18muaccbpm}
M. Otani et al., J. Phys. :Conf. Ser. 1067, 052012 (2018).

\bibitem{bib:Les75}
B. Leskovar and C. C. Lo, Nucl. Instrum. Meth. Phys. Res., Sect. A123, 145 (1975).

\bibitem{bib:Inami14}
K. Inami (Belle-II PID Group), Nucl. Instrum. Meth. Phys. Res., Sect. A766, 5 (2014).

\bibitem{bib:yotsuzuka19}
M. Yotsuzuka et al., Proceedings of IPAC2019, 2571 (2019). 

\bibitem{bib:Otani19buncher}
M.~Otani et al., Nucl. Inst. Meth. Phys. Res. Sec. A 946, 162693 (2019).

\bibitem{bib:sue20}
Y. Sue, M. Yotsuzuka et al., Phys. Rev. Accel. Beams 23, 022804 (2020).

\bibitem{bib:Ced21}
Ce Zhang et al., JPS Conf. Proc. 33, 011125 (2021). 

\bibitem{bib:Men20}
M. Lu et al.,  arXiv:physics.hep-ph/2010.15114 

\bibitem{bib:Kin21}
K. Cheung, Z. S. Wang., arXiv:physics.hep-ph/2101.10476

\end{thebibliography}

\end{document}